\documentclass[12pt,a4paper,aps,preprint,superscriptaddress,nofootinbib]{revtex4-1}

\usepackage{graphicx}
\usepackage{cancel}
\usepackage{amssymb}
\usepackage{textcomp}
\usepackage{amsmath}
\usepackage{bm}
\usepackage{times}
\usepackage{epsfig}
\usepackage{epstopdf}
\usepackage{color}
\usepackage{multirow}
\usepackage[colorlinks=true, pdfstartview=FitV, linkcolor=blue, citecolor=blue, urlcolor=blue]{hyperref}

\def\lsim{\mathrel{\raise.3ex\hbox{$<$\kern-.75em\lower1ex\hbox{$\sim$}}}}
\def\gsim{\mathrel{\raise.3ex\hbox{$>$\kern-.75em\lower1ex\hbox{$\sim$}}}}

\definecolor{orange}{rgb}{1,0.5,0}

\newcommand{\be}{\begin{equation}}
\newcommand{\ee}{\end{equation}}
\newcommand{\bea}{\begin{eqnarray}}
\newcommand{\eea}{\end{eqnarray}}

\begin{document}

\title{Simplified dark matter models with loop effects in direct detection and the constraints from indirect detection and collider search}

\author{Tong Li}
\email{litong@nankai.edu.cn}
\affiliation{
School of Physics, Nankai University, Tianjin 300071, China
}
\author{Peiwen Wu}
\email{pwwu@kias.re.kr}
\affiliation{
{School of Physics, KIAS, 85 Hoegiro, Seoul 02455, Republic of Korea}
}

\begin{abstract}
We reexamine the simplified dark matter (DM) models with fermionic DM particle and spin-0 mediator.
The DM-nucleon scattering cross sections of these models are low-momentum suppressed at tree-level, but receive sizable loop-induced spin-independent contribution. We perform one-loop calculation for scalar-type and twist-2 DM-quark operators and complete two-loop calculation for scalar-type DM-gluon operator. By analyzing the loop-level contribution from new operators, we find that future direct detection experiments can be sensitive to a fraction of parameter space. The indirect detection and collider search also provide complementary constraints on these models.
\end{abstract}


\maketitle

\section{Introduction}
\label{sec:Intro}

Although the existence of Dark Matter (DM) has been established by substantial cosmological and astronomical observations, the microscopic nature of DM particles is still unknown. An appealing candidate of DM is the Weakly Interacting Massive Particle (WIMP) arising from various extensions of the Standard Model (SM).
The experimental searches for WIMP consist of four main categories, i.e. the direct detection (DD) of possible scattering between DM and SM target materials, the indirect detection (ID) looking for signals of DM annihilation/decay products from the sky, the collider searches for signals from DM productions at high energy accelerators, and the gravitational and/or cosmological effects originating from the DM in the early and/or the current Universe.

Among the aforementioned four categories, the DD experiments have achieved significantly improved sensitivity in the past two decades, but yield null results up to now and very stringent bounds on the WIMP-nucleon scattering cross section.
A natural explanation of the absence of a confirmed DM signal is that the scattering rate is highly suppressed by the small typical value of transfer momentum of the process and/or the relative velocity between DM and nucleon. A simple but compelling scenario resulting in the suppressed rate at tree-level is that fermionic DM particles $\chi$ scatter off the target nucleon $N$ through a pseudo-scalar mediator in the $t$-channel scattering process~\cite{Boehm:2014hva,Ipek:2014gua}. The corresponding tree-level DM-nucleon contact interaction can reduce down to a non-relativistic contact operator
\begin{eqnarray}
\bar{\chi}i\gamma_5\chi N i\gamma_5 N \to (\mathbf{s}_\chi\cdot \mathbf{q})(\mathbf{s}_N\cdot \mathbf{q}),
\end{eqnarray}
in the non-relativistic limit.
Here, $\mathbf{s}_\chi$ ($\mathbf{s}_N$) is the DM (target nucleon) spin, and the scattering exchange momentum $\mathbf{q}$ is only of order 10 MeV. As a result, this scenario leads to a momentum-suppressed spin-dependent (SD) scattering cross section and thus an undetectable signal rate.

The suppression of tree-level scattering rate makes it appealing to further scrutinize the high-order effect from one-loop induced processes (see early discussions in e.g.~\cite{Drees:1993bu,Freytsis:2010ne}) and the possibly detectable signal at the upgrades of DD experiments. Integrating out the one-loop diagrams can induce distinct scalar-type operators giving non-momentum-suppressed spin-independent (SI) scattering cross section. The enhancement of loop-level SI cross section by the squared total nucleon number in a nucleus competes with the loop suppression and may dominate the WIMP-nuclei cross section over the suppressed tree-level scattering.
This one-loop effect in direct DM detection has been investigated in both simplified frameworks and UV complete models~\cite{Haisch:2013uaa,Ipek:2014gua,Arcadi:2017wqi,Sanderson:2018lmj,Li:2018qip,Han:2018gej,Abe:2018bpo,Ghorbani:2018pjh,Mohan:2019zrk}.
Recent progresses on the pseudo-scalar mediator scenario go beyond the one-loop processes for scalar-type DM-quark operator $m_q\bar{\chi}\chi \bar{q}q$ and include the dedicated contributions from two-loop scattering diagrams for scalar-type DM-gluon operator ${\alpha_s\over \pi}\bar{\chi}\chi GG$ after integrating out both heavy quarks and the mediator. It has been shown in Refs.~\cite{Abe:2018emu,Ertas:2019dew} that the full two-loop calculation deviates sizably from the result obtained by the conventional relation between the scalar-type current of heavy quarks and that of the gluon. This discrepancy is caused by the failure of the quark momentum expansion for heavy quarks and an ignored two-loop diagram for gluon emission when one utilized the relation for DM-gluon scattering.

In this work we revisit the loop effect in the DD of simplified DM models including but not limited to a pseudo-scalar mediator. We consider the hypotheses with one spin-0 mediator only coupled to the SM quarks and fermionic DM particles, giving momentum-suppressed WIMP-nuclei scattering cross sections at tree-level. The latest approaches of dedicated loop calculations are utilized in the high-order contributions to the cross sections, together with the estimate of the Renormalization Group Equations (RGE) running effects. We also take into account the constraints from other DM detection categories, e.g. the DM relic abundance, the ID constraint in terms of gamma-ray emission, as well as the current status of collider search. These synergistic discussions are regarded as a more complete improvement of the recently appeared works~\cite{Abe:2018emu,Ertas:2019dew}.

The paper is outlined as follows. In Sec.~\ref{sec:Model} we describe the simplified dark matter models. Then we give the effective DM-nucleon interactions at tree-level and the corresponding DM-nucleus scattering cross sections. In Sec.~\ref{sec:Loop}, we present the effective operators and Wilson coefficients at loop-level for the DM-nucleon cross section. The numerical results are given in Sec.~\ref{sec:Result}. Our conclusions are drawn in Sec.~\ref{sec:Con}. Some technical details for loop calculations are collected in the Appendix.
\section{Simplified dark matter hypothesis}
\label{sec:Model}

In this work, we focus on the simplified DM frameworks which consist of Majorana fermion DM $\chi$ and a spin-0 mediator $a$ coupled to $\chi$ and the SM quarks with strength $g_\chi$ and $g_q$ respectively. We consider each of the following three scenarios at a time
\begin{eqnarray}
&&\mathcal{L}_{\rm D2} = -{g_{\chi}\over 2} a \bar{\chi} i\gamma_5 \chi -g_q {m_q\over v_0} a \bar{q} q, \\
&&\mathcal{L}_{\rm D3} = -{g_{\chi}\over 2} a \bar{\chi} \chi -g_q {m_q\over v_0} a \bar{q} i\gamma_5 q, \\
&&\mathcal{L}_{\rm D4} = -{g_{\chi}\over 2} a \bar{\chi} i\gamma_5 \chi -g_q {m_q\over v_0} a \bar{q} i\gamma_5 q.
\end{eqnarray}
Here the $a\bar{q}q$ coupling is also scaled by the SM-like Yukawa coupling with $v_0 = 246$ GeV being the SM Higgs vacuum expectation value.
The model D4 with $a$ being a pure pseudo-scalar is designated as the pseudo-scalar mediator DM model in most of literatures. Models D2 and D3 are induced by more specific UV complete models with CP violation~\cite{LopezHonorez:2012kv,Beniwal:2015sdl,Baek:2017vzd,Athron:2018hpc,Abe:2019wku,Ertas:2019dew} and correspond to the cases with specific CP angles in Ref.~\cite{Ertas:2019dew}.

Based on the DM interactions with quarks and gluons at tree-level,
the DM-nucleon contact interactions are described by the effective Lagrangians as follows
\begin{eqnarray}
&&\mathcal{L}^{eff}_{\rm D2} = {C_N^{\rm tree}({\rm D2})\over 2m_{a}^2} \bar{\chi} i\gamma_5 \chi \bar{N} N, \ \mathcal{L}^{eff}_{\rm D3} = {C_N^{\rm tree}({\rm D_3})\over 2m_a^2} \bar{\chi} \chi \bar{N} i\gamma_5 N, \ \mathcal{L}^{eff}_{\rm D4} = {C_N^{\rm tree}({\rm D_4})\over 2m_a^2} \bar{\chi} i\gamma_5 \chi \bar{N} i\gamma_5 N, \nonumber \\
\end{eqnarray}
where the tree-level coefficients are defined as
\begin{eqnarray}
&&C_N^{\rm tree}({\rm D2}) = \sum_{q=u,d,s}{m_N\over m_q} C_q f_q^{N} + \sum_{q=c,b,t}{m_N\over m_q} C_q {2\over 27} f_G^{N}, \\
&&C_N^{\rm tree}({\rm D4}) = \sum_{q=u,d,s}{m_N\over m_q} \left(C_q-C\right) \Delta_q^{N}, \ \ \ C = \bar{m}\sum_{q=u,\cdots,t}{C_q\over m_q}, \ \ \ \bar{m}^{-1}=\sum_{q=u,d,s}m_q^{-1}.
\end{eqnarray}
The coefficient $C_N^{\rm tree}({\rm D_3})$ is equal to $C_N^{\rm tree}({\rm D_4})$ as models D3 and D4 share the same quark bilinear form $\bar{q}\gamma_5 q$.
Here $\Delta_q^{N}$, $f_q^{N}$ and $f_G^{N}$ are quark/gluon-nucleon form factors as numerically used in micrOMEGAs~\cite{Belanger:2013oya}.
The quark-level constant is defined as $C_q = g_\chi g_q{m_q\over v_0}$.
Consequently, their differential DM-nucleus scattering cross sections read as
\begin{eqnarray}
&&{d\sigma_{\rm SI}({\rm D2})\over dE_R} =
{1\over 32\pi} {m_T\over m_\chi^2 m_N^2 v^2} {4m_N^2q^2\over m_a^4} \sum_{N,N'=p,n}C_N^{\rm tree}({\rm D_2}) C_{N'}^{\rm tree}({\rm D_2}) F_{M}^{(N,N')}(q^2) ,
\label{D2xsectree}
\\
&&{d\sigma_{\rm SD}({\rm D3})\over dE_R} =
{1\over 32\pi} {m_T\over m_\chi^2 m_N^2 v^2} {4m_\chi^2 q^2\over m_a^4} \sum_{N,N'=p,n}C_N^{\rm tree}({\rm D_3}) C_{N'}^{\rm tree}({\rm D_3}) F_{\Sigma''}^{(N,N')}(q^2) ,
\label{D3xsectree}\\
&&{d\sigma_{\rm SD}({\rm D4})\over dE_R} =
{1\over 32\pi} {m_T\over m_\chi^2 m_N^2 v^2} {q^4\over m_a^4} \sum_{N,N'=p,n}C_N^{\rm tree}({\rm D_4}) C_{N'}^{\rm tree}({\rm D_4}) F_{\Sigma''}^{(N,N')}(q^2) ,
\label{D4xsectree}
\end{eqnarray}
where $m_T$ is the nucleus mass, $v$ is the DM speed in Earth's frame, $E_R$ is the nuclear recoil energy and $F_{\Sigma''}^{(N,N')}(q^2), F_{M}^{(N,N')}(q^2)$ are the form factors defined in Ref.~\cite{Fitzpatrick:2012ix}. The tree-level WIMP-nucleus scattering cross sections of the above simplified models are all dependent on the transfer momentum $q=\sqrt{2E_R m_T}$. As seen above, they are suppressed by $m_N^2q^2/ m_a^4$, $m_\chi^2 q^2/ m_a^4$ and $q^4/m_a^4$ for models D2, D3 and D4, respectively.

\section{Loop effect in direct detection}
\label{sec:Loop}
In this section, we derive the loop-level effect in direct DM detection of the above simplified models, followed by the estimate of the scale effects in terms of renormalization group evolutions.
\subsection{Loop effect from scalar-type quark/gluon operators}

\begin{figure}[h!]
\begin{center}
\includegraphics[scale=1,width=14cm]{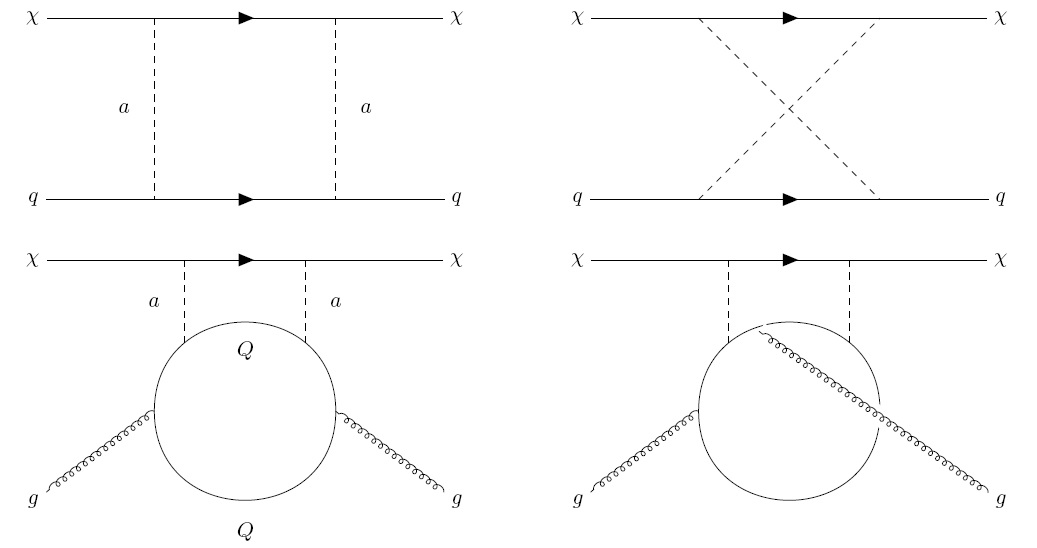}
\end{center}
\caption{Loop diagrams for the DM-quark currents (top) and DM-gluon current (bottom).
}
\label{loops}
\end{figure}

The general Lagrangian for the non-momentum-suppressed DM-nucleon SI cross section is given by
\begin{eqnarray}
\mathcal{L}_{eff} &=& {1\over 2}\sum_{q=u,d,s}C'_q m_q \bar{\chi}\chi \bar{q}q + {1\over 2}C_G\left(-{9\alpha_s\over 8\pi}\bar{\chi}\chi G^a_{\mu\nu}G^{a\mu\nu}\right)\nonumber \\
&+& {1\over 2}\sum_{q=u,d,s,c,b}\left[C_q^{(1)}\bar{\chi}i\partial^\mu \gamma^\nu \chi \mathcal{O}^q_{\mu\nu}+C_q^{(2)}\bar{\chi}i\partial^\mu i\partial^\nu \chi \mathcal{O}^q_{\mu\nu}\right],
\label{Lagrangian}
\end{eqnarray}
where $\mathcal{O}^q_{\mu\nu}={i\over 2}\bar{q}\left(\partial_\mu\gamma_\nu+\partial_\nu\gamma_\mu-{1\over 2}g_{\mu\nu}\cancel{\partial}\right)q$ is the twist-2 operator. For the models we consider, the Wilson coefficients in Eq.~(\ref{Lagrangian}) are all zero at tree-level but can be generated at loop-level, denoted by $C'_q=C_q^{\rm box}$, $C_G=C_G^{\rm box}$, $C_q^{(1)}=C_q^{\rm (1)box}$, $C_q^{(2)}=C_q^{\rm (2)box}$. The coefficients for scalar-type DM-quark operator and the twist-2 operator, i.e. $C_q^{\rm box}$, $C_q^{\rm (1)box}$ and $C_q^{\rm (2)box}$, are generated by the box diagrams in the top panels of Fig.~\ref{loops}. The two-loop diagrams in Fig.~\ref{loops} with only heavy quark $Q$ in the quark loop contribute to the scalar-type DM-gluon operator and the coefficient $C_G^{\rm box}$.

Following the non-relativistic limit used in Ref.~\cite{Abe:2018emu}, we expand the small momentum of valence quarks in the amplitude of the DM-quark scattering box diagrams. The coefficients $C_q^{\rm box}$, $C_q^{\rm (1)box}$ and $C_q^{\rm (2)box}$ are then obtained by reading out the DM-quark effective operators.
For the DM-gluon coefficient $C_G^{\rm box}$, one needs to calculate the amplitude of two-loop diagrams and find the effective operator $\bar{\chi}\chi G^a_{\mu\nu}G^{a\mu\nu}$. The complete two-loop calculations ensure the validity of the obtained $C_G^{\rm box}$ for any values of mediator mass $m_a$.
For model D4, the above Wilson coefficients are equivalent to those for pseudo-scalar mediator model as derived in Ref.~\cite{Abe:2018emu}
\begin{eqnarray}
C_q^{\rm box}({\rm D4})&=&{-m_\chi\over (4\pi)^2}\left({m_q\over v_0}\right)^2 {g_\chi^2 g_q^2\over m_a^2} [6X_{001}(m_\chi^2,m_\chi^2,0,m_a^2)+m_\chi^2X_{111}(m_\chi^2,m_\chi^2,0,m_a^2)\nonumber \\
&-&6X_{001}(m_\chi^2,m_\chi^2,m_a^2,0)-m_\chi^2X_{111}(m_\chi^2,m_\chi^2,m_a^2,0)],\\
C_q^{\rm (1)box}({\rm D4})&=&{-8\over (4\pi)^2}\left({m_q\over v_0}\right)^2 {g_\chi^2 g_q^2\over m_a^2}
[X_{001}(m_\chi^2,m_\chi^2,0,m_a^2)-X_{001}(m_\chi^2,m_\chi^2,m_a^2,0)],\\
C_q^{\rm (2)box}({\rm D4})&=&{-4m_\chi\over (4\pi)^2}\left({m_q\over v_0}\right)^2 {g_\chi^2 g_q^2\over m_a^2}
[X_{111}(m_\chi^2,m_\chi^2,0,m_a^2)-X_{111}(m_\chi^2,m_\chi^2,m_a^2,0)],\\
C_G^{\rm box}({\rm D4})&=&
\sum_{Q=c,b,t}{-m_\chi\over 432\pi^2}\left({m_Q\over v_0}\right)^2 g_\chi^2 g_q^2 {\partial F(m_a^2)\over \partial m_a^2},
\end{eqnarray}
where the loop functions $X_{001}, X_{111}, F$ are given in Ref.~\cite{Abe:2018emu} and the references therein.
Following the same procedure, for models D2 and D3, we obtain the corresponding Wilson coefficients which are related to those in model D4
\begin{eqnarray}
C_q^{\rm box}({\rm D2})&=&C_q^{\rm box}({\rm D4})\nonumber \\
&+&{4m_\chi\over (4\pi)^2}\left({m_q\over v_0}\right)^2 {g_\chi^2 g_q^2\over m_a^2}[C_2(m_\chi^2,m_a^2,m_\chi^2)+{1\over m_a^2}B_1(m_\chi^2,0,m_\chi^2)-{1\over m_a^2}B_1(m_\chi^2,m_a^2,m_\chi^2)],\nonumber \\
\\
C_q^{\rm box}({\rm D3})&=&C_q^{\rm box}({\rm D4})+{-8m_\chi\over (4\pi)^2}\left({m_q\over v_0}\right)^2 {g_\chi^2 g_q^2\over m_a^4}[X_{00}(m_\chi^2,m_a^2,m_\chi^2)+{m_\chi^2\over 4}X_{11}(m_\chi^2,m_a^2,m_\chi^2)],\\
C_q^{\rm (1)box}({\rm D2})&=&C_q^{\rm (1)box}({\rm D3})=C_q^{\rm (1)box}({\rm D4}), \ \ \ C_q^{\rm (2)box}({\rm D2})=C_q^{\rm (2)box}({\rm D4}),\\
C_q^{\rm (2)box}({\rm D3})&=&C_q^{\rm (2)box}({\rm D4})+{-8m_\chi\over (4\pi)^2}\left({m_q\over v_0}\right)^2 {g_\chi^2 g_q^2\over m_a^4}X_{11}(m_\chi^2,m_a^2,m_\chi^2),\\
C_G^{\rm box}({\rm D2})&=&\sum_{Q=c,b,t}{-m_\chi\over 432\pi^2}\left({m_Q\over v_0}\right)^2 g_\chi^2 g_q^2 {\partial F'(m_a^2)\over \partial m_a^2},\\
C_G^{\rm box}({\rm D3})&=&\sum_{Q=c,b,t}{-m_\chi\over 432\pi^2}\left({m_Q\over v_0}\right)^2 g_\chi^2 g_q^2 {\partial F''(m_a^2)\over \partial m_a^2}.
\end{eqnarray}
The new loop functions here are collected in Appendix. In our numerical calculation, we use Package-X~\cite{Patel:2015tea} to compute the above loop functions.

Based on the above effective operators for SI DM-nucleon scattering and the corresponding Wilson coefficients, we define the DM-nucleon constant at loop-level
\begin{eqnarray}
C_N^{\rm loop}&=&m_N\left[\sum_{q=u,d,s}C_q' f_{q}^N + C_G f_{G}^N + {3\over 4}\sum_{q=u,d,s,c,b}\left(m_\chi C_q^{(1)}+m_\chi^2C_q^{(2)}\right)\left(q^N(2)+\bar{q}^N(2)\right)\right],
\label{WCDMnucleon}
\end{eqnarray}
where the second moments of the parton distribution functions for quarks $q^N(2)$ and anti-quarks $\bar{q}^N(2)$ are taken from the CTEQ PDFs~\cite{Pumplin:2002vw}.
The SI cross section of the DM interaction with nucleon is thus given by
\begin{eqnarray}
\sigma_{\rm SI}={1\over \pi}\left({m_\chi m_N\over m_\chi + m_N}\right)^2 |C_N^{\rm loop}|^2.
\label{sigmaloop}
\end{eqnarray}
In terms of the form factor function $F_{M}^{(N,N')}(q^2)$, the differential SI cross section of the DM interaction with nucleus with mass $m_T$ is
\begin{eqnarray}
{d\sigma_{\rm SI}\over dE_R} =
{1\over 2\pi} {m_T\over v^2} \sum_{N,N'=p,n}C_N^{\rm loop} C_{N'}^{\rm loop} F_{M}^{(N,N')}(q^2).
\label{xsecloop}
\end{eqnarray}

\subsection{Loop effect from RGE running}

Another manifestation of loop effect is the mixing of operators according to the RGE. This RGE effect can be important if one considers DM phenomenology at vastly different energy scales. For instance, the DM annihilation typically happens at the electroweak scale and the DM particles are possibly produced near TeV scale at colliders. The energy scale for the DM-nucleon scattering in DD experiments is of the order of the hadron scale $\mu_{had}$. One usually starts with a gauge-invariant renormalizable DM model defined near or above the electroweak (EW) scale $\mu_{EW}\simeq m_Z$, but studies the non-relativistic DM-nucleon scattering rate with characteristic scale $\mu_{had}\simeq 1$ GeV. A series of effective field theories (EFTs) should be properly constructed by integrating out particles heavier than the current EFT scale $\mu_{EFT}$ and reasonably matched when passing thresholds of particles lighter than $\mu_{EFT}$ where they are integrated out in a similar way. Between the thresholds, the evolutions and mixings of the EFT operators should be performed according to RGE. The scale of the first EFT constructed in the whole analysis determines the procedures of RGE and threshold matching. The above procedures have been well elaborated in e.g.~\cite{Hisano:2010ct,Hisano:2015bma,Hisano:2015rsa,Hill:2014yka,Hill:2014yxa} and implemented in packages such as DirectDM~\cite{Bishara:2017nnn}, runDM~\cite{DEramo:2016gos}, Wilson~\cite{Aebischer:2018bkb} for specific or generic models.

An important difference should be emphasized between the scalar-type and twist-2 operators in Eq.~(\ref{Lagrangian}). The scalar-type form factors $f_q^{N}$ for light quarks $q=u,d,s$ are attributed to the non-perturbative QCD effects with energy scale around or below 1 GeV, and are obtained from the lattice QCD simulations. Thus, the scalar-type Wilson coefficients of light quarks $u,d,s$ and gluon must take values around 1 GeV.
Depending on the scale of the first EFT constructed in the whole analysis, e.g. at $\mu=m_Z$, this implies the procedures of RGE and threshold matching when calculating the scalar-type operator contributions.
At the scale of about 1 GeV, the heavy quarks $Q=c,b,t$ have been integrated out into the scalar-type gluon operator using the full two-loop calculations, as emphasized in Refs.~\cite{Abe:2018emu,Ertas:2019dew}. In turn, the twist-2 form factors $q^N(2), \bar{q}^N(2)$ in Eq.~(\ref{Lagrangian}) can be calculated perturbatively using parton PDFs at various scales \cite{Hisano:2010ct,Hisano:2015bma,Hisano:2015rsa,Hill:2014yka,Hill:2014yxa,Drees:1993bu}, e.g. 1 GeV or $m_Z$. One can choose a convenient scale to calculate the twist-2 contributions, with the proper active field contents (e.g. 5 flavor quarks $u,d,s,c,b$ at $\mu=m_Z$) and the Wilson coefficients and form factors evaluated at that scale.
Note that we ignored the negligible contributions from twist-2 gluon operator, since its Wilson coefficient is suppressed by an additional $\alpha_s/\pi$ due to the operator definition~\cite{Hisano:2015bma,Hisano:2015rsa}.

We note that in Refs.~\cite{Abe:2018emu,Ertas:2019dew}, the values of $q^N(2), \bar{q}^N(2)$ are evaluated at $\mu=m_Z$ in the calculations.
Since the same (similar) box diagrams and model parameters are used to obtain the scalar-type and twist-2 Wilson coefficients for quarks (gluons), a more consistent implementation should involve the RGE running effects for the scale-type operators from $m_Z$ to 1 GeV as discussed above. Since the coupling between the mediator and the SM quarks is chosen to mimic the SM Yukawa structure, the DM-nucleon constant from scalar-type DM-gluon interaction $C_G f_{G}^N$ in Eq.~(\ref{WCDMnucleon}) is dominated by the top quark loop. The constant $C_G f_{G}^N$ also dominates over the scalar-type and twist-2 DM-quark interactions. To have a conservative estimate of the scale effects on the scalar type DM-gluon operator $-{9\alpha_s\over 8\pi}\bar{\chi}\chi G^a_{\mu\nu}G^{a\mu\nu}$, we utilize the package DirectDM to perform its RGE running from $m_Z$ to 1 GeV. We find that the scale effects give a negative correction of $1\%\sim 2\%$, and thus do not affect our main conclusions in this work.

\section{Other dark matter constraints}
In this section, we consider DM constraints on the above simplified models from other categories mentioned in the Introduction, including the relic abundance, indirect detection and collider search.

Assuming that DM particles have frozen out in the early Universe, as standard thermal relics, they acquire their present abundance through annihilation processes.
The pairs of DM particle $\chi$ in the simplified models can either annihilate into SM quark or gluon pairs via $s$-channel processes $\chi \chi\to a\to q\bar{q}, gg$ or annihilate into two mediators $\chi \chi\to a a$ when kinematically allowed~\cite{Arina:2014yna,Abdallah:2015ter,Boveia:2016mrp,Balazs:2017hxh,Li:2017nac}. The amplitudes of the two annihilation channels are governed by $g_\chi g_q$ and $g_\chi^2$, respectively. We assume all kinematically accessible final states of the DM annihilation and use micrOMEGAs 5.0~\cite{Belanger:2018mqt} to calculate the relic abundance. Note that the WIMP candidate here may account only a fraction of the total DM of the Universe, referred as multi-component DM scenario~\cite{Zurek:2008qg,Profumo:2009tb}. In this scenario, the DM energy density measured by PLANCK~\cite{Ade:2015xua} is imposed as an upper limit on the WIMP relic abundance.

Dwarf galaxies are the search targets for DM annihilation into gamma rays.
The Fermi Large Area Telescope (LAT) has detected no excess of gamma ray emission from the dwarf spheroidal satellite galaxies (dSphs) of the Milky Way. Thus, an upper limit on the DM annihilation cross section can be placed from a combined analysis of multiple Milky Way dSphs~\cite{Ackermann:2015zua,Fermi-LAT:2016uux}.
For individual dwarf galaxy target, Fermi-LAT provided tabulated values of delta-log-likelihood as a function of the energy flux bin-by-bin.
The gamma ray energy flux from DM annihilation for the $j$th energy bin and the $k$th dwarf is given by
\begin{eqnarray}
\Phi^E_{j,k}(m_{\chi},\langle \sigma v\rangle,J_k)=\frac{\langle \sigma v\rangle}{16\pi m_{\chi}^2}J_k\int^{E^{\rm max}_j}_{E^{\rm min}_j}E\frac{dN_\gamma}{dE}dE,
\end{eqnarray}
where $J_k$ is the J factor for the $k$th dwarf. The energy flux only depends on $m_{\chi}$, $\langle \sigma v\rangle$ and $J_k$, and is thus calculable for the DM annihilation process from the above simplified models. We use the PPPC4DMID package~\cite{Cirelli:2010xx} to obtain the spectrum of photons $dN_\gamma/dE$. The likelihood for $k$th dwarf is
\begin{eqnarray}
&&\mathcal{L}_k(m_{\chi},\langle \sigma v\rangle,J_k)=\mathcal{L}_J(J_k|\bar{J}_k,\sigma_k)\prod_j \mathcal{L}_{j,k}(\Phi^E_{j,k}(m_{\chi},\langle \sigma v\rangle,J_k)),
\end{eqnarray}
where $\mathcal{L}_{j,k}$ is the tabulated likelihood provided by Fermi-LAT for each dwarf and energy flux. The uncertainty of the J factors is taken into account by profiling over $J_k$ in the likelihood below~\cite{Ackermann:2015zua}
\begin{eqnarray}
&&\mathcal{L}_J(J_k|\bar{J}_k,\sigma_k)={1\over \ln(10)J_k\sqrt{2\pi}\sigma_k}\times e^{-(\log_{10}(J_k)-\log_{10}(\bar{J}_k))^2/2\sigma_k^2},
\end{eqnarray}
with the measured J factor $\bar{J}_k$ and error $\sigma_k$.
A joint likelihood for all dwarfs can then be performed as
\begin{eqnarray}
\mathcal{L}(m_{\chi},\langle \sigma v\rangle,\mathbb{J})=\prod_k \mathcal{L}_k(m_{\chi},\langle \sigma v\rangle,J_k),
\end{eqnarray}
where $\mathbb{J}$ is the set of J factors $J_k$. In our numerical implementation, we adopt the corresponding values of $\mathcal{L}_{j,k}$ and $\bar{J}_k, \sigma_k$ for 19 dwarf galaxies considered in Ref.~\cite{Fermi-LAT:2016uux}.

According to the maximum likelihood analysis adopted by Fermi-LAT, the delta-log-likelihood is given by
\begin{eqnarray}
-2\Delta \ln \mathcal{L}(m_{\chi},\langle \sigma v\rangle)=-2\ln\left({\mathcal{L}(m_{\chi},\langle \sigma v\rangle,\widehat{\widehat{\mathbb{J}}})\over \mathcal{L}(m_{\chi},\widehat{\langle \sigma v\rangle},\widehat{\mathbb{J}})}\right),
\end{eqnarray}
where $\widehat{\langle \sigma v\rangle}$ and $\widehat{\mathbb{J}}$ maximize the likelihood at any given $m_{\chi}$, and $\widehat{\widehat{\mathbb{J}}}$ maximizes the likelihood for given $m_{\chi}$ and $\langle \sigma v\rangle$.
The 95\% C.L. upper limit on the annihilation cross section for a given $m_{\chi}$ is determined by demanding $-2\Delta \ln\mathcal{L}(m_{\chi},\langle \sigma v\rangle)\leq 2.71$. We perform the likelihood analysis and obtain the upper limit using Minuit~\cite{James:1975dr}.
Once the annihilation cross section calculated by a certain set of model parameters is larger than the limit, we claim the corresponding parameter values are excluded by Fermi-LAT dSphs.

The Large Hadron Collider (LHC) performed the search for DM in association with energetic jet~\cite{Aaboud:2017phn,Sirunyan:2017jix} or the third generation quarks~\cite{Aaboud:2017rzf,Aaboud:2017aeu,Sirunyan:2017xgm} for simplified DM models with spin-0 mediator at $\sqrt{s}=13$ TeV collisions. The most severe limits are from final states with $t\bar{t}$ and missing transverse momentum~\cite{Aaboud:2019yqu}.
For model D4 with pseudo-scalar mediator, assuming unitary couplings $g_\chi=g_q=1$, the range of mediator mass between 15 and 25 GeV is excluded~\cite{Aaboud:2017aeu}.
This limit is valid for all DM masses as long as $m_a>2m_\chi$ and closely related models D2 and D3 should have very similar collider constraint.

\section{Results}
\label{sec:Result}

\begin{figure}[h!]
\begin{center}
\includegraphics[scale=1,width=7cm]{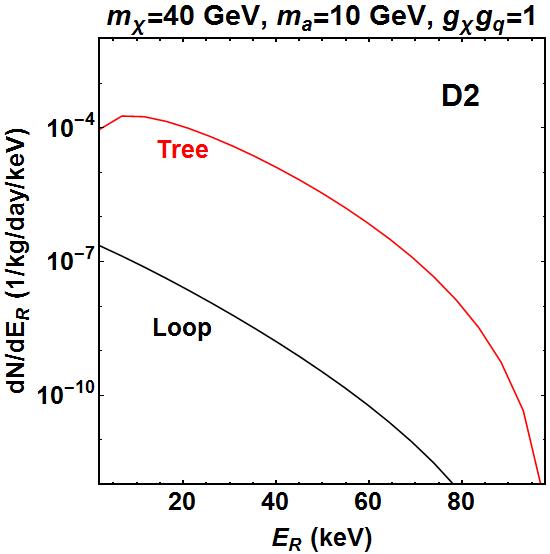}
\includegraphics[scale=1,width=7cm]{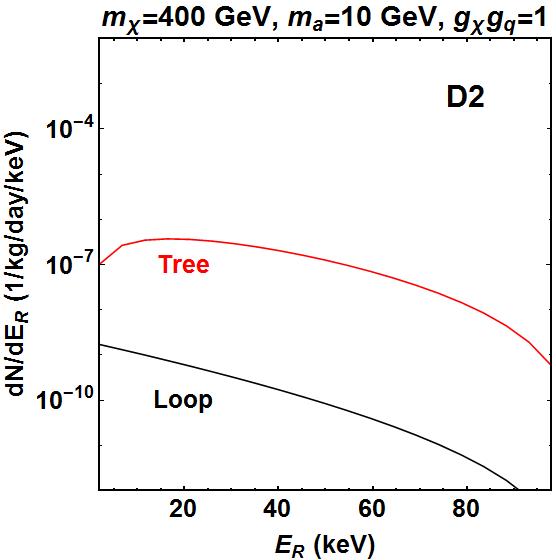}
\end{center}
\caption{
Recoil energy spectra of model D2 with $m_a=10$ GeV, $g_\chi g_q=1$ and $m_\chi = 40$ GeV (left) or $m_\chi = 400$ GeV (right).
The tree-level and loop-level spectra are denoted by red and black curves, respectively.
}
\label{dNdE_D2}
\end{figure}

\begin{figure}[h!]
\begin{center}
\includegraphics[scale=1,width=7cm]{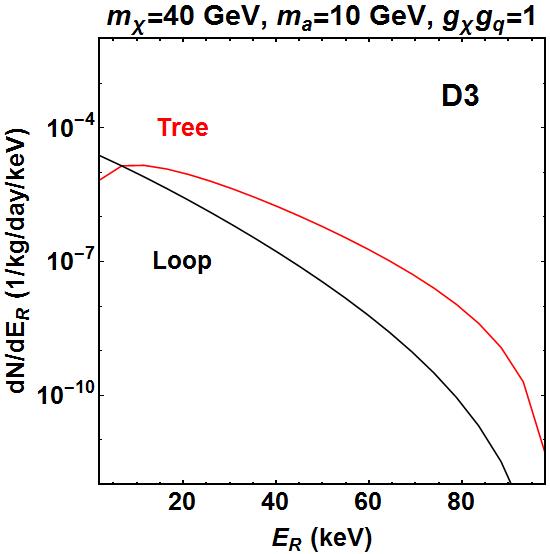}
\includegraphics[scale=1,width=7cm]{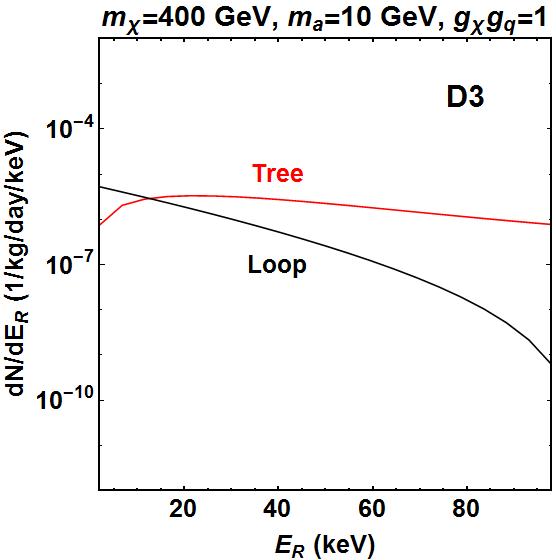}
\end{center}
\caption{
Recoil energy spectra of model D3, as labeled in Fig.~\ref{dNdE_D2}.
}
\label{dNdE_D3}
\end{figure}

\begin{figure}[h!]
\begin{center}
\includegraphics[scale=1,width=7cm]{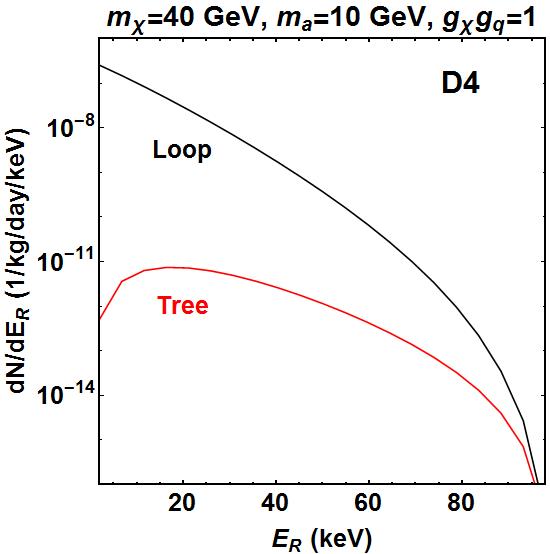}
\includegraphics[scale=1,width=7cm]{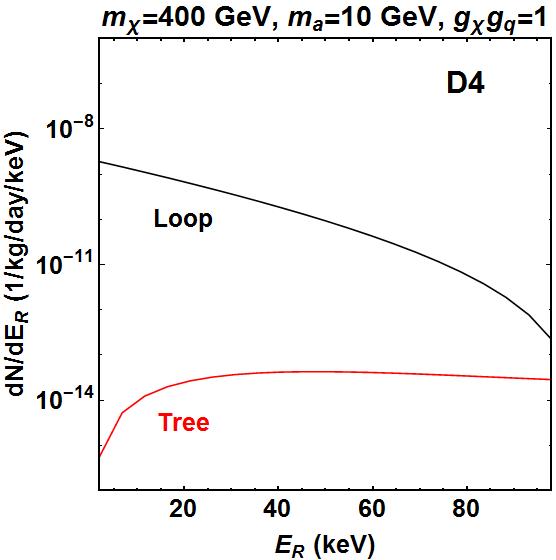}
\end{center}
\caption{Recoil energy spectra of model D4, as labeled in Fig.~\ref{dNdE_D2}.
}
\label{dNdE_D4}
\end{figure}

By combining the theoretical calculations of DM-nucleus scattering cross sections with a certain velocity distribution for DM particles, we can calculate WIMP signal rates for DD experiments.
The differential event rate with respect to the recoil energy is given by
\begin{eqnarray}
{dN\over dE_R}={\rho_\chi\over m_\chi}\int d^3v vf(\vec{v}){d\sigma(v,E_R)\over dE_R},
\end{eqnarray}
where $\rho_\chi$ is the local DM density which is fixed to be $0.3 \ {\rm GeV}/{\rm cm}^3$. We take a cut-off Maxwell-Boltzmann velocity distribution for $f(\vec{v})$ with the escape velocity as $v_{esc}=544 \ {\rm km}/{\rm s}$.
Together with the differential scattering cross sections obtained above, the differential event rate can be evaluated and in practice we employ DMFormFactor~\cite{Fitzpatrick:2012ix,Anand:2013yka} for the numerical calculation on xenon nucleus $^{129}{\rm Xe}$.
The recoil energy spectra of models D2, D3 and D4 are shown in Figs.~\ref{dNdE_D2}, \ref{dNdE_D3} and \ref{dNdE_D4}, respectively. Model D2 leads to SI scattering cross section with a strong enhancement for large nuclei. Thus, although there is momentum suppression at tree-level, the spectrum at tree-level for model D2 still dominates over the loop-level contribution. As being suppressed by $q^4$ at tree-level, in turn, the loop-level spectrum is much greater than that at tree-level for model D4. Although models D3 and D4 both lead to SD cross section at tree-level, this discrepancy is smaller in model D3 as its tree-level scattering cross section is suppressed by $q^2$ only. The loop-level spectrum of model D3 dominates over the tree-level one only in the range of low recoil energy.

\begin{figure}[h!]
\begin{center}
\includegraphics[scale=1,width=8cm]{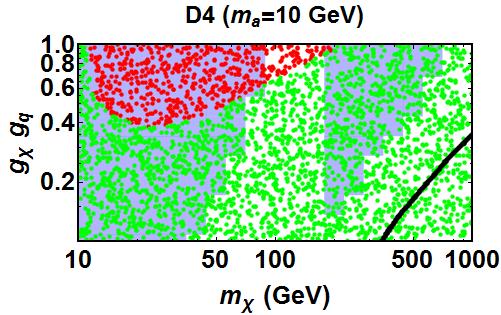}
\includegraphics[scale=1,width=8cm]{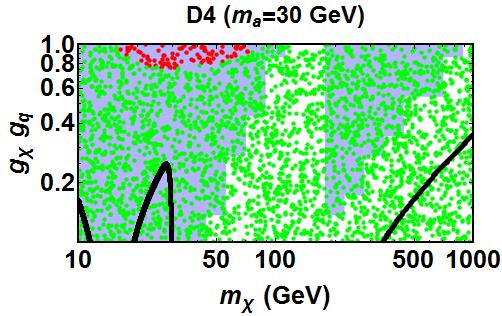}
\end{center}
\caption{The region of $g_\chi g_q$ vs. $m_\chi$ below the Xenon1T exclusion limit and above the neutrino floor (red) and the region below the neutrino floor (green) for model D4, with $m_a=10$ GeV (left) and $m_a=30$ GeV (right). Assuming $g_\chi = 1$, the light blue values are excluded by Fermi-LAT dSphs and the DM relic abundance satisfies $\Omega h^2\leq 0.12$ above the black curves.
}
\label{D4-g}
\end{figure}

As the recoil spectrum of the SI scattering induced by loop diagrams is dominant in model D4, the prediction of SI DM-nucleon cross section in Eq.~(\ref{sigmaloop}) can be compared directly to the limits set by DD experiments to yield a bound on $g_\chi g_q$. As shown in Fig.~\ref{D4-g}, for $g_\chi g_q \leq 1$ in model D4, the SI scattering cross sections in the red region are below the XENON1T exclusion limit~\cite{Aprile:2017iyp,Aprile:2018dbl} but above the neutrino floor. The green region gives cross sections below the neutrino floor. Future DD experiments can thus be more sensitive to the region of $g_\chi g_q > 0.4$ and $m_\chi < 200$ GeV in the case of $m_a=10$ GeV.

The cross section of DM annihilation into SM quarks for model D4 is proportional to $m_q^2/v_0^2$, thus the $t\bar{t}$ channel dominates if kinematically allowed. The annihilation to mediator pairs is governed by $m_a/m_\chi$ and plays a crucial role in small $m_\chi$ range.
The Fermi-LAT dSphs exclude a majority of parameter region for $m_{\chi}\lesssim 100$ GeV and $m_{\chi}\gtrsim m_t$, as shown in light blue region in Fig.~\ref{D4-g}. The region around $m_{\chi}\simeq 100$ GeV evades the ID constraint due to the fact that the $\chi\chi\to a\to t\bar{t}$ channel is not kinematically allowed and the annihilation to mediator pairs is suppressed~\cite{Balazs:2017hxh,Li:2017nac}.
Above the black contours in Fig.~\ref{D4-g}, the DM relic abundance satisfies $\Omega h^2\leq 0.12$.

\section{Conclusion}
\label{sec:Con}

We reexamined the loop-level correction to the WIMP-nucleon scattering cross section in the framework of simplified DM models. A spin-0 mediator is assumed to couple with fermionic DM particle and the SM quarks in each model. The cross sections of these models are low-momentum suppressed at tree-level, but receive sizable loop-induced SI contribution. Following the recent progress on the loop-level correction, we perform one-loop calculation for scalar-type and twist-2 DM-quark operators and complete two-loop calculation for scalar-type DM-gluon operator. By including the loop-level SI cross section, we find that future DD experiments can be sensitive to a fraction of parameter space which gives no detectable signal with only tree-level contribution. The sensitivity of DD experiments to these models is also complementary to the constraints from ID and collider search.

\acknowledgments
T.~L. would like to thank Tomohiro Abe, Motoko Fujiwara and Junji Hisano for helpful discussion. P.~W. would like to thank Chengcheng Han for the beneficial argument.
The work of T.~L. is supported by ``the Fundamental Research Funds for the Central Universities'', Nankai University (Grant Number 63196013).

\appendix
\section{Loop functions}

The used loop functions for models D2 and D3 are
\begin{eqnarray}
\int {d^D \ell \over (2\pi)^D} {\ell_\mu \ell_\nu \over [(\ell+p)^2-m_\chi^2](\ell^2-m_a^2)^2\ell^4} = {i\over (4\pi)^2 m_a^4} \left[ g_{\mu\nu} X_{00}(p^2,m_a^2,m_\chi^2)+p_\mu p_\nu X_{11}(p^2,m_a^2,m_\chi^2)\right] \nonumber \\
\end{eqnarray}
and
\begin{eqnarray}
&&F'(m_a^2)=\int^1_0 dx [3Y_1(m_\chi^2,m_\chi^2,m_a^2,m_Q^2)-m_Q^2{-x^2-3x\over x^2(1-x)^2}Y_2(m_\chi^2,m_\chi^2,m_a^2,m_Q^2)\nonumber \\
&&+4m_Q^4{x^2(1-2x)\over x^3(1-x)^3}Y_3(m_\chi^2,m_\chi^2,m_a^2,m_Q^2)],\\
&&F''(m_a^2)=\int^1_0 dx [3Y_1(m_\chi^2,m_\chi^2,m_a^2,m_Q^2)-m_Q^2{9x-5x^2\over x^2(1-x)^2}Y_2(m_\chi^2,m_\chi^2,m_a^2,m_Q^2)\nonumber \\
&&-2m_Q^4{2x^2\over x^3(1-x)^3}Y_3(m_\chi^2,m_\chi^2,m_a^2,m_Q^2)]\nonumber \\
&&+2\int^1_0 dx [3Z_1(m_\chi^2,m_\chi^2,m_a^2,m_Q^2)-m_Q^2{9x-5x^2\over x^2(1-x)^2}Z_2(m_\chi^2,m_\chi^2,m_a^2,m_Q^2)\nonumber \\
&&-2m_Q^4{2x^2\over x^3(1-x)^3}Z_3(m_\chi^2,m_\chi^2,m_a^2,m_Q^2)],
\end{eqnarray}
with
\begin{eqnarray}
&&\int {d^D\ell\over (2\pi)^D}{1\over [(\ell+p)^2-m_\chi^2](\ell^2-m_a^2)[\ell^2-{m_Q^2\over x(1-x)}]}={i\over (4\pi)^2}Z_1(p^2,m_\chi^2,m_a^2,m_Q^2),\\
&&\int {d^D\ell\over (2\pi)^D}{1\over [(\ell+p)^2-m_\chi^2](\ell^2-m_a^2)[\ell^2-{m_Q^2\over x(1-x)}]^2}={i\over (4\pi)^2}Z_2(p^2,m_\chi^2,m_a^2,m_Q^2),\\
&&\int {d^D\ell\over (2\pi)^D}{1\over [(\ell+p)^2-m_\chi^2](\ell^2-m_a^2)[\ell^2-{m_Q^2\over x(1-x)}]^3}={i\over (4\pi)^2}Z_3(p^2,m_\chi^2,m_a^2,m_Q^2).
\end{eqnarray}
Here the $Y_i (i=1,2,3)$ functions are shown in Ref.~\cite{Abe:2018emu} and the references therein.

\bibliography{refs}

\begin{thebibliography}{53}%
\makeatletter
\providecommand \@ifxundefined [1]{%
 \@ifx{#1\undefined}
}%
\providecommand \@ifnum [1]{%
 \ifnum #1\expandafter \@firstoftwo
 \else \expandafter \@secondoftwo
 \fi
}%
\providecommand \@ifx [1]{%
 \ifx #1\expandafter \@firstoftwo
 \else \expandafter \@secondoftwo
 \fi
}%
\providecommand \natexlab [1]{#1}%
\providecommand \enquote  [1]{``#1''}%
\providecommand \bibnamefont  [1]{#1}%
\providecommand \bibfnamefont [1]{#1}%
\providecommand \citenamefont [1]{#1}%
\providecommand \href@noop [0]{\@secondoftwo}%
\providecommand \href [0]{\begingroup \@sanitize@url \@href}%
\providecommand \@href[1]{\@@startlink{#1}\@@href}%
\providecommand \@@href[1]{\endgroup#1\@@endlink}%
\providecommand \@sanitize@url [0]{\catcode `\\12\catcode `\$12\catcode
  `\&12\catcode `\#12\catcode `\^12\catcode `\_12\catcode `\%12\relax}%
\providecommand \@@startlink[1]{}%
\providecommand \@@endlink[0]{}%
\providecommand \url  [0]{\begingroup\@sanitize@url \@url }%
\providecommand \@url [1]{\endgroup\@href {#1}{\urlprefix }}%
\providecommand \urlprefix  [0]{URL }%
\providecommand \Eprint [0]{\href }%
\providecommand \doibase [0]{http://dx.doi.org/}%
\providecommand \selectlanguage [0]{\@gobble}%
\providecommand \bibinfo  [0]{\@secondoftwo}%
\providecommand \bibfield  [0]{\@secondoftwo}%
\providecommand \translation [1]{[#1]}%
\providecommand \BibitemOpen [0]{}%
\providecommand \bibitemStop [0]{}%
\providecommand \bibitemNoStop [0]{.\EOS\space}%
\providecommand \EOS [0]{\spacefactor3000\relax}%
\providecommand \BibitemShut  [1]{\csname bibitem#1\endcsname}%
\let\auto@bib@innerbib\@empty
\bibitem [{\citenamefont {Boehm}\ \emph {et~al.}(2014)\citenamefont {Boehm},
  \citenamefont {Dolan}, \citenamefont {McCabe}, \citenamefont {Spannowsky},\
  and\ \citenamefont {Wallace}}]{Boehm:2014hva}%
  \BibitemOpen
  \bibfield  {author} {\bibinfo {author} {\bibfnamefont {C.}~\bibnamefont
  {Boehm}}, \bibinfo {author} {\bibfnamefont {M.~J.}\ \bibnamefont {Dolan}},
  \bibinfo {author} {\bibfnamefont {C.}~\bibnamefont {McCabe}}, \bibinfo
  {author} {\bibfnamefont {M.}~\bibnamefont {Spannowsky}}, \ and\ \bibinfo
  {author} {\bibfnamefont {C.~J.}\ \bibnamefont {Wallace}},\ }\href {\doibase
  10.1088/1475-7516/2014/05/009} {\bibfield  {journal} {\bibinfo  {journal}
  {JCAP}\ }\textbf {\bibinfo {volume} {1405}},\ \bibinfo {pages} {009}
  (\bibinfo {year} {2014})},\ \Eprint {http://arxiv.org/abs/1401.6458}
  {arXiv:1401.6458 [hep-ph]} \BibitemShut {NoStop}%
\bibitem [{\citenamefont {Ipek}\ \emph {et~al.}(2014)\citenamefont {Ipek},
  \citenamefont {McKeen},\ and\ \citenamefont {Nelson}}]{Ipek:2014gua}%
  \BibitemOpen
  \bibfield  {author} {\bibinfo {author} {\bibfnamefont {S.}~\bibnamefont
  {Ipek}}, \bibinfo {author} {\bibfnamefont {D.}~\bibnamefont {McKeen}}, \ and\
  \bibinfo {author} {\bibfnamefont {A.~E.}\ \bibnamefont {Nelson}},\ }\href
  {\doibase 10.1103/PhysRevD.90.055021} {\bibfield  {journal} {\bibinfo
  {journal} {Phys. Rev.}\ }\textbf {\bibinfo {volume} {D90}},\ \bibinfo {pages}
  {055021} (\bibinfo {year} {2014})},\ \Eprint {http://arxiv.org/abs/1404.3716}
  {arXiv:1404.3716 [hep-ph]} \BibitemShut {NoStop}%
\bibitem [{\citenamefont {Drees}\ and\ \citenamefont
  {Nojiri}(1993)}]{Drees:1993bu}%
  \BibitemOpen
  \bibfield  {author} {\bibinfo {author} {\bibfnamefont {M.}~\bibnamefont
  {Drees}}\ and\ \bibinfo {author} {\bibfnamefont {M.}~\bibnamefont {Nojiri}},\
  }\href {\doibase 10.1103/PhysRevD.48.3483} {\bibfield  {journal} {\bibinfo
  {journal} {Phys. Rev.}\ }\textbf {\bibinfo {volume} {D48}},\ \bibinfo {pages}
  {3483} (\bibinfo {year} {1993})},\ \Eprint
  {http://arxiv.org/abs/hep-ph/9307208} {arXiv:hep-ph/9307208 [hep-ph]}
  \BibitemShut {NoStop}%
\bibitem [{\citenamefont {Freytsis}\ and\ \citenamefont
  {Ligeti}(2011)}]{Freytsis:2010ne}%
  \BibitemOpen
  \bibfield  {author} {\bibinfo {author} {\bibfnamefont {M.}~\bibnamefont
  {Freytsis}}\ and\ \bibinfo {author} {\bibfnamefont {Z.}~\bibnamefont
  {Ligeti}},\ }\href {\doibase 10.1103/PhysRevD.83.115009} {\bibfield
  {journal} {\bibinfo  {journal} {Phys. Rev.}\ }\textbf {\bibinfo {volume}
  {D83}},\ \bibinfo {pages} {115009} (\bibinfo {year} {2011})},\ \Eprint
  {http://arxiv.org/abs/1012.5317} {arXiv:1012.5317 [hep-ph]} \BibitemShut
  {NoStop}%
\bibitem [{\citenamefont {Haisch}\ and\ \citenamefont
  {Kahlhoefer}(2013)}]{Haisch:2013uaa}%
  \BibitemOpen
  \bibfield  {author} {\bibinfo {author} {\bibfnamefont {U.}~\bibnamefont
  {Haisch}}\ and\ \bibinfo {author} {\bibfnamefont {F.}~\bibnamefont
  {Kahlhoefer}},\ }\href {\doibase 10.1088/1475-7516/2013/04/050} {\bibfield
  {journal} {\bibinfo  {journal} {JCAP}\ }\textbf {\bibinfo {volume} {1304}},\
  \bibinfo {pages} {050} (\bibinfo {year} {2013})},\ \Eprint
  {http://arxiv.org/abs/1302.4454} {arXiv:1302.4454 [hep-ph]} \BibitemShut
  {NoStop}%
\bibitem [{\citenamefont {Arcadi}\ \emph {et~al.}(2018)\citenamefont {Arcadi},
  \citenamefont {Lindner}, \citenamefont {Queiroz}, \citenamefont
  {Rodejohann},\ and\ \citenamefont {Vogl}}]{Arcadi:2017wqi}%
  \BibitemOpen
  \bibfield  {author} {\bibinfo {author} {\bibfnamefont {G.}~\bibnamefont
  {Arcadi}}, \bibinfo {author} {\bibfnamefont {M.}~\bibnamefont {Lindner}},
  \bibinfo {author} {\bibfnamefont {F.~S.}\ \bibnamefont {Queiroz}}, \bibinfo
  {author} {\bibfnamefont {W.}~\bibnamefont {Rodejohann}}, \ and\ \bibinfo
  {author} {\bibfnamefont {S.}~\bibnamefont {Vogl}},\ }\href {\doibase
  10.1088/1475-7516/2018/03/042} {\bibfield  {journal} {\bibinfo  {journal}
  {JCAP}\ }\textbf {\bibinfo {volume} {1803}},\ \bibinfo {pages} {042}
  (\bibinfo {year} {2018})},\ \Eprint {http://arxiv.org/abs/1711.02110}
  {arXiv:1711.02110 [hep-ph]} \BibitemShut {NoStop}%
\bibitem [{\citenamefont {Bell}\ \emph {et~al.}(2018)\citenamefont {Bell},
  \citenamefont {Busoni},\ and\ \citenamefont {Sanderson}}]{Sanderson:2018lmj}%
  \BibitemOpen
  \bibfield  {author} {\bibinfo {author} {\bibfnamefont {N.~F.}\ \bibnamefont
  {Bell}}, \bibinfo {author} {\bibfnamefont {G.}~\bibnamefont {Busoni}}, \ and\
  \bibinfo {author} {\bibfnamefont {I.~W.}\ \bibnamefont {Sanderson}},\ }\href
  {\doibase 10.1088/1475-7516/2018/08/017, 10.1088/1475-7516/2019/01/E01}
  {\bibfield  {journal} {\bibinfo  {journal} {JCAP}\ }\textbf {\bibinfo
  {volume} {1808}},\ \bibinfo {pages} {017} (\bibinfo {year} {2018})},\
  \bibinfo {note} {[Erratum: JCAP1901,no.01,E01(2019)]},\ \Eprint
  {http://arxiv.org/abs/1803.01574} {arXiv:1803.01574 [hep-ph]} \BibitemShut
  {NoStop}%
\bibitem [{\citenamefont {Li}(2018{\natexlab{a}})}]{Li:2018qip}%
  \BibitemOpen
  \bibfield  {author} {\bibinfo {author} {\bibfnamefont {T.}~\bibnamefont
  {Li}},\ }\href {\doibase 10.1016/j.physletb.2018.05.073} {\bibfield
  {journal} {\bibinfo  {journal} {Phys. Lett.}\ }\textbf {\bibinfo {volume}
  {B782}},\ \bibinfo {pages} {497} (\bibinfo {year} {2018}{\natexlab{a}})},\
  \Eprint {http://arxiv.org/abs/1804.02120} {arXiv:1804.02120 [hep-ph]}
  \BibitemShut {NoStop}%
\bibitem [{\citenamefont {Han}\ \emph {et~al.}(2019)\citenamefont {Han},
  \citenamefont {Liu}, \citenamefont {Mukhopadhyay},\ and\ \citenamefont
  {Wang}}]{Han:2018gej}%
  \BibitemOpen
  \bibfield  {author} {\bibinfo {author} {\bibfnamefont {T.}~\bibnamefont
  {Han}}, \bibinfo {author} {\bibfnamefont {H.}~\bibnamefont {Liu}}, \bibinfo
  {author} {\bibfnamefont {S.}~\bibnamefont {Mukhopadhyay}}, \ and\ \bibinfo
  {author} {\bibfnamefont {X.}~\bibnamefont {Wang}},\ }\href {\doibase
  10.1007/JHEP03(2019)080} {\bibfield  {journal} {\bibinfo  {journal} {JHEP}\
  }\textbf {\bibinfo {volume} {03}},\ \bibinfo {pages} {080} (\bibinfo {year}
  {2019})},\ \Eprint {http://arxiv.org/abs/1810.04679} {arXiv:1810.04679
  [hep-ph]} \BibitemShut {NoStop}%
\bibitem [{\citenamefont {Abe}\ \emph {et~al.}(2018)\citenamefont {Abe} \emph
  {et~al.}}]{Abe:2018bpo}%
  \BibitemOpen
  \bibfield  {author} {\bibinfo {author} {\bibfnamefont {T.}~\bibnamefont
  {Abe}} \emph {et~al.} (\bibinfo {collaboration} {LHC Dark Matter Working
  Group}),\ }\href@noop {} {\  (\bibinfo {year} {2018})},\ \Eprint
  {http://arxiv.org/abs/1810.09420} {arXiv:1810.09420 [hep-ex]} \BibitemShut
  {NoStop}%
\bibitem [{\citenamefont {Ghorbani}\ and\ \citenamefont
  {Ghorbani}(2018)}]{Ghorbani:2018pjh}%
  \BibitemOpen
  \bibfield  {author} {\bibinfo {author} {\bibfnamefont {K.}~\bibnamefont
  {Ghorbani}}\ and\ \bibinfo {author} {\bibfnamefont {P.~H.}\ \bibnamefont
  {Ghorbani}},\ }\href@noop {} {\  (\bibinfo {year} {2018})},\ \Eprint
  {http://arxiv.org/abs/1812.04092} {arXiv:1812.04092 [hep-ph]} \BibitemShut
  {NoStop}%
\bibitem [{\citenamefont {Mohan}\ \emph {et~al.}(2019)\citenamefont {Mohan},
  \citenamefont {Sengupta}, \citenamefont {Tait}, \citenamefont {Yan},\ and\
  \citenamefont {Yuan}}]{Mohan:2019zrk}%
  \BibitemOpen
  \bibfield  {author} {\bibinfo {author} {\bibfnamefont {K.~A.}\ \bibnamefont
  {Mohan}}, \bibinfo {author} {\bibfnamefont {D.}~\bibnamefont {Sengupta}},
  \bibinfo {author} {\bibfnamefont {T.~M.~P.}\ \bibnamefont {Tait}}, \bibinfo
  {author} {\bibfnamefont {B.}~\bibnamefont {Yan}}, \ and\ \bibinfo {author}
  {\bibfnamefont {P.}~\bibnamefont {Yuan}},\ }\href@noop {} {\  (\bibinfo
  {year} {2019})},\ \Eprint {http://arxiv.org/abs/1903.05650} {arXiv:1903.05650
  [hep-ph]} \BibitemShut {NoStop}%
\bibitem [{\citenamefont {Abe}\ \emph {et~al.}(2019)\citenamefont {Abe},
  \citenamefont {Fujiwara},\ and\ \citenamefont {Hisano}}]{Abe:2018emu}%
  \BibitemOpen
  \bibfield  {author} {\bibinfo {author} {\bibfnamefont {T.}~\bibnamefont
  {Abe}}, \bibinfo {author} {\bibfnamefont {M.}~\bibnamefont {Fujiwara}}, \
  and\ \bibinfo {author} {\bibfnamefont {J.}~\bibnamefont {Hisano}},\ }\href
  {\doibase 10.1007/JHEP02(2019)028} {\bibfield  {journal} {\bibinfo  {journal}
  {JHEP}\ }\textbf {\bibinfo {volume} {02}},\ \bibinfo {pages} {028} (\bibinfo
  {year} {2019})},\ \Eprint {http://arxiv.org/abs/1810.01039} {arXiv:1810.01039
  [hep-ph]} \BibitemShut {NoStop}%
\bibitem [{\citenamefont {Ertas}\ and\ \citenamefont
  {Kahlhoefer}(2019)}]{Ertas:2019dew}%
  \BibitemOpen
  \bibfield  {author} {\bibinfo {author} {\bibfnamefont {F.}~\bibnamefont
  {Ertas}}\ and\ \bibinfo {author} {\bibfnamefont {F.}~\bibnamefont
  {Kahlhoefer}},\ }\href@noop {} {\  (\bibinfo {year} {2019})},\ \Eprint
  {http://arxiv.org/abs/1902.11070} {arXiv:1902.11070 [hep-ph]} \BibitemShut
  {NoStop}%
\bibitem [{\citenamefont {Lopez-Honorez}\ \emph {et~al.}(2012)\citenamefont
  {Lopez-Honorez}, \citenamefont {Schwetz},\ and\ \citenamefont
  {Zupan}}]{LopezHonorez:2012kv}%
  \BibitemOpen
  \bibfield  {author} {\bibinfo {author} {\bibfnamefont {L.}~\bibnamefont
  {Lopez-Honorez}}, \bibinfo {author} {\bibfnamefont {T.}~\bibnamefont
  {Schwetz}}, \ and\ \bibinfo {author} {\bibfnamefont {J.}~\bibnamefont
  {Zupan}},\ }\href {\doibase 10.1016/j.physletb.2012.07.017} {\bibfield
  {journal} {\bibinfo  {journal} {Phys. Lett.}\ }\textbf {\bibinfo {volume}
  {B716}},\ \bibinfo {pages} {179} (\bibinfo {year} {2012})},\ \Eprint
  {http://arxiv.org/abs/1203.2064} {arXiv:1203.2064 [hep-ph]} \BibitemShut
  {NoStop}%
\bibitem [{\citenamefont {Beniwal}\ \emph {et~al.}(2016)\citenamefont
  {Beniwal}, \citenamefont {Rajec}, \citenamefont {Savage}, \citenamefont
  {Scott}, \citenamefont {Weniger}, \citenamefont {White},\ and\ \citenamefont
  {Williams}}]{Beniwal:2015sdl}%
  \BibitemOpen
  \bibfield  {author} {\bibinfo {author} {\bibfnamefont {A.}~\bibnamefont
  {Beniwal}}, \bibinfo {author} {\bibfnamefont {F.}~\bibnamefont {Rajec}},
  \bibinfo {author} {\bibfnamefont {C.}~\bibnamefont {Savage}}, \bibinfo
  {author} {\bibfnamefont {P.}~\bibnamefont {Scott}}, \bibinfo {author}
  {\bibfnamefont {C.}~\bibnamefont {Weniger}}, \bibinfo {author} {\bibfnamefont
  {M.}~\bibnamefont {White}}, \ and\ \bibinfo {author} {\bibfnamefont {A.~G.}\
  \bibnamefont {Williams}},\ }\href {\doibase 10.1103/PhysRevD.93.115016}
  {\bibfield  {journal} {\bibinfo  {journal} {Phys. Rev.}\ }\textbf {\bibinfo
  {volume} {D93}},\ \bibinfo {pages} {115016} (\bibinfo {year} {2016})},\
  \Eprint {http://arxiv.org/abs/1512.06458} {arXiv:1512.06458 [hep-ph]}
  \BibitemShut {NoStop}%
\bibitem [{\citenamefont {Baek}\ \emph {et~al.}(2017)\citenamefont {Baek},
  \citenamefont {Ko},\ and\ \citenamefont {Li}}]{Baek:2017vzd}%
  \BibitemOpen
  \bibfield  {author} {\bibinfo {author} {\bibfnamefont {S.}~\bibnamefont
  {Baek}}, \bibinfo {author} {\bibfnamefont {P.}~\bibnamefont {Ko}}, \ and\
  \bibinfo {author} {\bibfnamefont {J.}~\bibnamefont {Li}},\ }\href {\doibase
  10.1103/PhysRevD.95.075011} {\bibfield  {journal} {\bibinfo  {journal} {Phys.
  Rev.}\ }\textbf {\bibinfo {volume} {D95}},\ \bibinfo {pages} {075011}
  (\bibinfo {year} {2017})},\ \Eprint {http://arxiv.org/abs/1701.04131}
  {arXiv:1701.04131 [hep-ph]} \BibitemShut {NoStop}%
\bibitem [{\citenamefont {Athron}\ \emph {et~al.}(2019)\citenamefont {Athron}
  \emph {et~al.}}]{Athron:2018hpc}%
  \BibitemOpen
  \bibfield  {author} {\bibinfo {author} {\bibfnamefont {P.}~\bibnamefont
  {Athron}} \emph {et~al.} (\bibinfo {collaboration} {GAMBIT}),\ }\href
  {\doibase 10.1140/epjc/s10052-018-6513-6} {\bibfield  {journal} {\bibinfo
  {journal} {Eur. Phys. J.}\ }\textbf {\bibinfo {volume} {C79}},\ \bibinfo
  {pages} {38} (\bibinfo {year} {2019})},\ \Eprint
  {http://arxiv.org/abs/1808.10465} {arXiv:1808.10465 [hep-ph]} \BibitemShut
  {NoStop}%
\bibitem [{\citenamefont {Abe}\ and\ \citenamefont {Sato}(2019)}]{Abe:2019wku}%
  \BibitemOpen
  \bibfield  {author} {\bibinfo {author} {\bibfnamefont {T.}~\bibnamefont
  {Abe}}\ and\ \bibinfo {author} {\bibfnamefont {R.}~\bibnamefont {Sato}},\
  }\href {\doibase 10.1103/PhysRevD.99.035012} {\bibfield  {journal} {\bibinfo
  {journal} {Phys. Rev.}\ }\textbf {\bibinfo {volume} {D99}},\ \bibinfo {pages}
  {035012} (\bibinfo {year} {2019})},\ \Eprint
  {http://arxiv.org/abs/1901.02278} {arXiv:1901.02278 [hep-ph]} \BibitemShut
  {NoStop}%
\bibitem [{\citenamefont {Belanger}\ \emph {et~al.}(2014)\citenamefont
  {Belanger}, \citenamefont {Boudjema}, \citenamefont {Pukhov},\ and\
  \citenamefont {Semenov}}]{Belanger:2013oya}%
  \BibitemOpen
  \bibfield  {author} {\bibinfo {author} {\bibfnamefont {G.}~\bibnamefont
  {Belanger}}, \bibinfo {author} {\bibfnamefont {F.}~\bibnamefont {Boudjema}},
  \bibinfo {author} {\bibfnamefont {A.}~\bibnamefont {Pukhov}}, \ and\ \bibinfo
  {author} {\bibfnamefont {A.}~\bibnamefont {Semenov}},\ }\href {\doibase
  10.1016/j.cpc.2013.10.016} {\bibfield  {journal} {\bibinfo  {journal}
  {Comput. Phys. Commun.}\ }\textbf {\bibinfo {volume} {185}},\ \bibinfo
  {pages} {960} (\bibinfo {year} {2014})},\ \Eprint
  {http://arxiv.org/abs/1305.0237} {arXiv:1305.0237 [hep-ph]} \BibitemShut
  {NoStop}%
\bibitem [{\citenamefont {Fitzpatrick}\ \emph {et~al.}(2013)\citenamefont
  {Fitzpatrick}, \citenamefont {Haxton}, \citenamefont {Katz}, \citenamefont
  {Lubbers},\ and\ \citenamefont {Xu}}]{Fitzpatrick:2012ix}%
  \BibitemOpen
  \bibfield  {author} {\bibinfo {author} {\bibfnamefont {A.~L.}\ \bibnamefont
  {Fitzpatrick}}, \bibinfo {author} {\bibfnamefont {W.}~\bibnamefont {Haxton}},
  \bibinfo {author} {\bibfnamefont {E.}~\bibnamefont {Katz}}, \bibinfo {author}
  {\bibfnamefont {N.}~\bibnamefont {Lubbers}}, \ and\ \bibinfo {author}
  {\bibfnamefont {Y.}~\bibnamefont {Xu}},\ }\href {\doibase
  10.1088/1475-7516/2013/02/004} {\bibfield  {journal} {\bibinfo  {journal}
  {JCAP}\ }\textbf {\bibinfo {volume} {1302}},\ \bibinfo {pages} {004}
  (\bibinfo {year} {2013})},\ \Eprint {http://arxiv.org/abs/1203.3542}
  {arXiv:1203.3542 [hep-ph]} \BibitemShut {NoStop}%
\bibitem [{\citenamefont {Patel}(2015)}]{Patel:2015tea}%
  \BibitemOpen
  \bibfield  {author} {\bibinfo {author} {\bibfnamefont {H.~H.}\ \bibnamefont
  {Patel}},\ }\href {\doibase 10.1016/j.cpc.2015.08.017} {\bibfield  {journal}
  {\bibinfo  {journal} {Comput. Phys. Commun.}\ }\textbf {\bibinfo {volume}
  {197}},\ \bibinfo {pages} {276} (\bibinfo {year} {2015})},\ \Eprint
  {http://arxiv.org/abs/1503.01469} {arXiv:1503.01469 [hep-ph]} \BibitemShut
  {NoStop}%
\bibitem [{\citenamefont {Pumplin}\ \emph {et~al.}(2002)\citenamefont
  {Pumplin}, \citenamefont {Stump}, \citenamefont {Huston}, \citenamefont
  {Lai}, \citenamefont {Nadolsky},\ and\ \citenamefont
  {Tung}}]{Pumplin:2002vw}%
  \BibitemOpen
  \bibfield  {author} {\bibinfo {author} {\bibfnamefont {J.}~\bibnamefont
  {Pumplin}}, \bibinfo {author} {\bibfnamefont {D.~R.}\ \bibnamefont {Stump}},
  \bibinfo {author} {\bibfnamefont {J.}~\bibnamefont {Huston}}, \bibinfo
  {author} {\bibfnamefont {H.~L.}\ \bibnamefont {Lai}}, \bibinfo {author}
  {\bibfnamefont {P.~M.}\ \bibnamefont {Nadolsky}}, \ and\ \bibinfo {author}
  {\bibfnamefont {W.~K.}\ \bibnamefont {Tung}},\ }\href {\doibase
  10.1088/1126-6708/2002/07/012} {\bibfield  {journal} {\bibinfo  {journal}
  {JHEP}\ }\textbf {\bibinfo {volume} {07}},\ \bibinfo {pages} {012} (\bibinfo
  {year} {2002})},\ \Eprint {http://arxiv.org/abs/hep-ph/0201195}
  {arXiv:hep-ph/0201195 [hep-ph]} \BibitemShut {NoStop}%
\bibitem [{\citenamefont {Hisano}\ \emph {et~al.}(2010)\citenamefont {Hisano},
  \citenamefont {Ishiwata},\ and\ \citenamefont {Nagata}}]{Hisano:2010ct}%
  \BibitemOpen
  \bibfield  {author} {\bibinfo {author} {\bibfnamefont {J.}~\bibnamefont
  {Hisano}}, \bibinfo {author} {\bibfnamefont {K.}~\bibnamefont {Ishiwata}}, \
  and\ \bibinfo {author} {\bibfnamefont {N.}~\bibnamefont {Nagata}},\ }\href
  {\doibase 10.1103/PhysRevD.82.115007} {\bibfield  {journal} {\bibinfo
  {journal} {Phys. Rev.}\ }\textbf {\bibinfo {volume} {D82}},\ \bibinfo {pages}
  {115007} (\bibinfo {year} {2010})},\ \Eprint {http://arxiv.org/abs/1007.2601}
  {arXiv:1007.2601 [hep-ph]} \BibitemShut {NoStop}%
\bibitem [{\citenamefont {Hisano}\ \emph
  {et~al.}(2015{\natexlab{a}})\citenamefont {Hisano}, \citenamefont {Nagai},\
  and\ \citenamefont {Nagata}}]{Hisano:2015bma}%
  \BibitemOpen
  \bibfield  {author} {\bibinfo {author} {\bibfnamefont {J.}~\bibnamefont
  {Hisano}}, \bibinfo {author} {\bibfnamefont {R.}~\bibnamefont {Nagai}}, \
  and\ \bibinfo {author} {\bibfnamefont {N.}~\bibnamefont {Nagata}},\ }\href
  {\doibase 10.1007/JHEP05(2015)037} {\bibfield  {journal} {\bibinfo  {journal}
  {JHEP}\ }\textbf {\bibinfo {volume} {05}},\ \bibinfo {pages} {037} (\bibinfo
  {year} {2015}{\natexlab{a}})},\ \Eprint {http://arxiv.org/abs/1502.02244}
  {arXiv:1502.02244 [hep-ph]} \BibitemShut {NoStop}%
\bibitem [{\citenamefont {Hisano}\ \emph
  {et~al.}(2015{\natexlab{b}})\citenamefont {Hisano}, \citenamefont
  {Ishiwata},\ and\ \citenamefont {Nagata}}]{Hisano:2015rsa}%
  \BibitemOpen
  \bibfield  {author} {\bibinfo {author} {\bibfnamefont {J.}~\bibnamefont
  {Hisano}}, \bibinfo {author} {\bibfnamefont {K.}~\bibnamefont {Ishiwata}}, \
  and\ \bibinfo {author} {\bibfnamefont {N.}~\bibnamefont {Nagata}},\ }\href
  {\doibase 10.1007/JHEP06(2015)097} {\bibfield  {journal} {\bibinfo  {journal}
  {JHEP}\ }\textbf {\bibinfo {volume} {06}},\ \bibinfo {pages} {097} (\bibinfo
  {year} {2015}{\natexlab{b}})},\ \Eprint {http://arxiv.org/abs/1504.00915}
  {arXiv:1504.00915 [hep-ph]} \BibitemShut {NoStop}%
\bibitem [{\citenamefont {Hill}\ and\ \citenamefont
  {Solon}(2015{\natexlab{a}})}]{Hill:2014yka}%
  \BibitemOpen
  \bibfield  {author} {\bibinfo {author} {\bibfnamefont {R.~J.}\ \bibnamefont
  {Hill}}\ and\ \bibinfo {author} {\bibfnamefont {M.~P.}\ \bibnamefont
  {Solon}},\ }\href {\doibase 10.1103/PhysRevD.91.043504} {\bibfield  {journal}
  {\bibinfo  {journal} {Phys. Rev.}\ }\textbf {\bibinfo {volume} {D91}},\
  \bibinfo {pages} {043504} (\bibinfo {year} {2015}{\natexlab{a}})},\ \Eprint
  {http://arxiv.org/abs/1401.3339} {arXiv:1401.3339 [hep-ph]} \BibitemShut
  {NoStop}%
\bibitem [{\citenamefont {Hill}\ and\ \citenamefont
  {Solon}(2015{\natexlab{b}})}]{Hill:2014yxa}%
  \BibitemOpen
  \bibfield  {author} {\bibinfo {author} {\bibfnamefont {R.~J.}\ \bibnamefont
  {Hill}}\ and\ \bibinfo {author} {\bibfnamefont {M.~P.}\ \bibnamefont
  {Solon}},\ }\href {\doibase 10.1103/PhysRevD.91.043505} {\bibfield  {journal}
  {\bibinfo  {journal} {Phys. Rev.}\ }\textbf {\bibinfo {volume} {D91}},\
  \bibinfo {pages} {043505} (\bibinfo {year} {2015}{\natexlab{b}})},\ \Eprint
  {http://arxiv.org/abs/1409.8290} {arXiv:1409.8290 [hep-ph]} \BibitemShut
  {NoStop}%
\bibitem [{\citenamefont {Bishara}\ \emph {et~al.}(2017)\citenamefont
  {Bishara}, \citenamefont {Brod}, \citenamefont {Grinstein},\ and\
  \citenamefont {Zupan}}]{Bishara:2017nnn}%
  \BibitemOpen
  \bibfield  {author} {\bibinfo {author} {\bibfnamefont {F.}~\bibnamefont
  {Bishara}}, \bibinfo {author} {\bibfnamefont {J.}~\bibnamefont {Brod}},
  \bibinfo {author} {\bibfnamefont {B.}~\bibnamefont {Grinstein}}, \ and\
  \bibinfo {author} {\bibfnamefont {J.}~\bibnamefont {Zupan}},\ }\href@noop {}
  {\  (\bibinfo {year} {2017})},\ \Eprint {http://arxiv.org/abs/1708.02678}
  {arXiv:1708.02678 [hep-ph]} \BibitemShut {NoStop}%
\bibitem [{\citenamefont {D'Eramo}\ \emph {et~al.}(2016)\citenamefont
  {D'Eramo}, \citenamefont {Kavanagh},\ and\ \citenamefont
  {Panci}}]{DEramo:2016gos}%
  \BibitemOpen
  \bibfield  {author} {\bibinfo {author} {\bibfnamefont {F.}~\bibnamefont
  {D'Eramo}}, \bibinfo {author} {\bibfnamefont {B.~J.}\ \bibnamefont
  {Kavanagh}}, \ and\ \bibinfo {author} {\bibfnamefont {P.}~\bibnamefont
  {Panci}},\ }\href {\doibase 10.1007/JHEP08(2016)111} {\bibfield  {journal}
  {\bibinfo  {journal} {JHEP}\ }\textbf {\bibinfo {volume} {08}},\ \bibinfo
  {pages} {111} (\bibinfo {year} {2016})},\ \Eprint
  {http://arxiv.org/abs/1605.04917} {arXiv:1605.04917 [hep-ph]} \BibitemShut
  {NoStop}%
\bibitem [{\citenamefont {Aebischer}\ \emph {et~al.}(2018)\citenamefont
  {Aebischer}, \citenamefont {Kumar},\ and\ \citenamefont
  {Straub}}]{Aebischer:2018bkb}%
  \BibitemOpen
  \bibfield  {author} {\bibinfo {author} {\bibfnamefont {J.}~\bibnamefont
  {Aebischer}}, \bibinfo {author} {\bibfnamefont {J.}~\bibnamefont {Kumar}}, \
  and\ \bibinfo {author} {\bibfnamefont {D.~M.}\ \bibnamefont {Straub}},\
  }\href {\doibase 10.1140/epjc/s10052-018-6492-7} {\bibfield  {journal}
  {\bibinfo  {journal} {Eur. Phys. J.}\ }\textbf {\bibinfo {volume} {C78}},\
  \bibinfo {pages} {1026} (\bibinfo {year} {2018})},\ \Eprint
  {http://arxiv.org/abs/1804.05033} {arXiv:1804.05033 [hep-ph]} \BibitemShut
  {NoStop}%
\bibitem [{\citenamefont {Arina}\ \emph {et~al.}(2015)\citenamefont {Arina},
  \citenamefont {Del~Nobile},\ and\ \citenamefont {Panci}}]{Arina:2014yna}%
  \BibitemOpen
  \bibfield  {author} {\bibinfo {author} {\bibfnamefont {C.}~\bibnamefont
  {Arina}}, \bibinfo {author} {\bibfnamefont {E.}~\bibnamefont {Del~Nobile}}, \
  and\ \bibinfo {author} {\bibfnamefont {P.}~\bibnamefont {Panci}},\ }\href
  {\doibase 10.1103/PhysRevLett.114.011301} {\bibfield  {journal} {\bibinfo
  {journal} {Phys. Rev. Lett.}\ }\textbf {\bibinfo {volume} {114}},\ \bibinfo
  {pages} {011301} (\bibinfo {year} {2015})},\ \Eprint
  {http://arxiv.org/abs/1406.5542} {arXiv:1406.5542 [hep-ph]} \BibitemShut
  {NoStop}%
\bibitem [{\citenamefont {Abdallah}\ \emph {et~al.}(2015)\citenamefont
  {Abdallah} \emph {et~al.}}]{Abdallah:2015ter}%
  \BibitemOpen
  \bibfield  {author} {\bibinfo {author} {\bibfnamefont {J.}~\bibnamefont
  {Abdallah}} \emph {et~al.},\ }\href {\doibase 10.1016/j.dark.2015.08.001}
  {\bibfield  {journal} {\bibinfo  {journal} {Phys. Dark Univ.}\ }\textbf
  {\bibinfo {volume} {9-10}},\ \bibinfo {pages} {8} (\bibinfo {year} {2015})},\
  \Eprint {http://arxiv.org/abs/1506.03116} {arXiv:1506.03116 [hep-ph]}
  \BibitemShut {NoStop}%
\bibitem [{\citenamefont {Busoni}\ \emph {et~al.}(2016)\citenamefont {Busoni}
  \emph {et~al.}}]{Boveia:2016mrp}%
  \BibitemOpen
  \bibfield  {author} {\bibinfo {author} {\bibfnamefont {G.}~\bibnamefont
  {Busoni}} \emph {et~al.},\ }\href@noop {} {\  (\bibinfo {year} {2016})},\
  \Eprint {http://arxiv.org/abs/1603.04156} {arXiv:1603.04156 [hep-ex]}
  \BibitemShut {NoStop}%
\bibitem [{\citenamefont {Bal\'{a}zs}\ \emph {et~al.}(2017)\citenamefont
  {Bal\'{a}zs}, \citenamefont {Conrad}, \citenamefont {Farmer}, \citenamefont
  {Jacques}, \citenamefont {Li}, \citenamefont {Meyer}, \citenamefont
  {Queiroz},\ and\ \citenamefont {S\'{a}nchez-Conde}}]{Balazs:2017hxh}%
  \BibitemOpen
  \bibfield  {author} {\bibinfo {author} {\bibfnamefont {C.}~\bibnamefont
  {Bal\'{a}zs}}, \bibinfo {author} {\bibfnamefont {J.}~\bibnamefont {Conrad}},
  \bibinfo {author} {\bibfnamefont {B.}~\bibnamefont {Farmer}}, \bibinfo
  {author} {\bibfnamefont {T.}~\bibnamefont {Jacques}}, \bibinfo {author}
  {\bibfnamefont {T.}~\bibnamefont {Li}}, \bibinfo {author} {\bibfnamefont
  {M.}~\bibnamefont {Meyer}}, \bibinfo {author} {\bibfnamefont {F.~S.}\
  \bibnamefont {Queiroz}}, \ and\ \bibinfo {author} {\bibfnamefont {M.~A.}\
  \bibnamefont {S\'{a}nchez-Conde}},\ }\href {\doibase
  10.1103/PhysRevD.96.083002} {\bibfield  {journal} {\bibinfo  {journal} {Phys.
  Rev.}\ }\textbf {\bibinfo {volume} {D96}},\ \bibinfo {pages} {083002}
  (\bibinfo {year} {2017})},\ \Eprint {http://arxiv.org/abs/1706.01505}
  {arXiv:1706.01505 [astro-ph.HE]} \BibitemShut {NoStop}%
\bibitem [{\citenamefont {Li}(2018{\natexlab{b}})}]{Li:2017nac}%
  \BibitemOpen
  \bibfield  {author} {\bibinfo {author} {\bibfnamefont {T.}~\bibnamefont
  {Li}},\ }\href {\doibase 10.1007/JHEP01(2018)151} {\bibfield  {journal}
  {\bibinfo  {journal} {JHEP}\ }\textbf {\bibinfo {volume} {01}},\ \bibinfo
  {pages} {151} (\bibinfo {year} {2018}{\natexlab{b}})},\ \Eprint
  {http://arxiv.org/abs/1708.04534} {arXiv:1708.04534 [hep-ph]} \BibitemShut
  {NoStop}%
\bibitem [{\citenamefont {B\'{e}langer}\ \emph {et~al.}(2018)\citenamefont
  {B\'{e}langer}, \citenamefont {Boudjema}, \citenamefont {Goudelis},
  \citenamefont {Pukhov},\ and\ \citenamefont {Zaldivar}}]{Belanger:2018mqt}%
  \BibitemOpen
  \bibfield  {author} {\bibinfo {author} {\bibfnamefont {G.}~\bibnamefont
  {B\'{e}langer}}, \bibinfo {author} {\bibfnamefont {F.}~\bibnamefont
  {Boudjema}}, \bibinfo {author} {\bibfnamefont {A.}~\bibnamefont {Goudelis}},
  \bibinfo {author} {\bibfnamefont {A.}~\bibnamefont {Pukhov}}, \ and\ \bibinfo
  {author} {\bibfnamefont {B.}~\bibnamefont {Zaldivar}},\ }\href {\doibase
  10.1016/j.cpc.2018.04.027} {\bibfield  {journal} {\bibinfo  {journal}
  {Comput. Phys. Commun.}\ }\textbf {\bibinfo {volume} {231}},\ \bibinfo
  {pages} {173} (\bibinfo {year} {2018})},\ \Eprint
  {http://arxiv.org/abs/1801.03509} {arXiv:1801.03509 [hep-ph]} \BibitemShut
  {NoStop}%
\bibitem [{\citenamefont {Zurek}(2009)}]{Zurek:2008qg}%
  \BibitemOpen
  \bibfield  {author} {\bibinfo {author} {\bibfnamefont {K.~M.}\ \bibnamefont
  {Zurek}},\ }\href {\doibase 10.1103/PhysRevD.79.115002} {\bibfield  {journal}
  {\bibinfo  {journal} {Phys. Rev.}\ }\textbf {\bibinfo {volume} {D79}},\
  \bibinfo {pages} {115002} (\bibinfo {year} {2009})},\ \Eprint
  {http://arxiv.org/abs/0811.4429} {arXiv:0811.4429 [hep-ph]} \BibitemShut
  {NoStop}%
\bibitem [{\citenamefont {Profumo}\ \emph {et~al.}(2009)\citenamefont
  {Profumo}, \citenamefont {Sigurdson},\ and\ \citenamefont
  {Ubaldi}}]{Profumo:2009tb}%
  \BibitemOpen
  \bibfield  {author} {\bibinfo {author} {\bibfnamefont {S.}~\bibnamefont
  {Profumo}}, \bibinfo {author} {\bibfnamefont {K.}~\bibnamefont {Sigurdson}},
  \ and\ \bibinfo {author} {\bibfnamefont {L.}~\bibnamefont {Ubaldi}},\ }\href
  {\doibase 10.1088/1475-7516/2009/12/016} {\bibfield  {journal} {\bibinfo
  {journal} {JCAP}\ }\textbf {\bibinfo {volume} {0912}},\ \bibinfo {pages}
  {016} (\bibinfo {year} {2009})},\ \Eprint {http://arxiv.org/abs/0907.4374}
  {arXiv:0907.4374 [hep-ph]} \BibitemShut {NoStop}%
\bibitem [{\citenamefont {Ade}\ \emph {et~al.}(2016)\citenamefont {Ade} \emph
  {et~al.}}]{Ade:2015xua}%
  \BibitemOpen
  \bibfield  {author} {\bibinfo {author} {\bibfnamefont {P.~A.~R.}\
  \bibnamefont {Ade}} \emph {et~al.} (\bibinfo {collaboration} {Planck}),\
  }\href {\doibase 10.1051/0004-6361/201525830} {\bibfield  {journal} {\bibinfo
   {journal} {Astron. Astrophys.}\ }\textbf {\bibinfo {volume} {594}},\
  \bibinfo {pages} {A13} (\bibinfo {year} {2016})},\ \Eprint
  {http://arxiv.org/abs/1502.01589} {arXiv:1502.01589 [astro-ph.CO]}
  \BibitemShut {NoStop}%
\bibitem [{\citenamefont {Ackermann}\ \emph {et~al.}(2015)\citenamefont
  {Ackermann} \emph {et~al.}}]{Ackermann:2015zua}%
  \BibitemOpen
  \bibfield  {author} {\bibinfo {author} {\bibfnamefont {M.}~\bibnamefont
  {Ackermann}} \emph {et~al.} (\bibinfo {collaboration} {Fermi-LAT}),\ }\href
  {\doibase 10.1103/PhysRevLett.115.231301} {\bibfield  {journal} {\bibinfo
  {journal} {Phys. Rev. Lett.}\ }\textbf {\bibinfo {volume} {115}},\ \bibinfo
  {pages} {231301} (\bibinfo {year} {2015})},\ \Eprint
  {http://arxiv.org/abs/1503.02641} {arXiv:1503.02641 [astro-ph.HE]}
  \BibitemShut {NoStop}%
\bibitem [{\citenamefont {Albert}\ \emph {et~al.}(2017)\citenamefont {Albert}
  \emph {et~al.}}]{Fermi-LAT:2016uux}%
  \BibitemOpen
  \bibfield  {author} {\bibinfo {author} {\bibfnamefont {A.}~\bibnamefont
  {Albert}} \emph {et~al.} (\bibinfo {collaboration} {Fermi-LAT, DES}),\ }\href
  {\doibase 10.3847/1538-4357/834/2/110} {\bibfield  {journal} {\bibinfo
  {journal} {Astrophys. J.}\ }\textbf {\bibinfo {volume} {834}},\ \bibinfo
  {pages} {110} (\bibinfo {year} {2017})},\ \Eprint
  {http://arxiv.org/abs/1611.03184} {arXiv:1611.03184 [astro-ph.HE]}
  \BibitemShut {NoStop}%
\bibitem [{\citenamefont {Cirelli}\ \emph {et~al.}(2011)\citenamefont
  {Cirelli}, \citenamefont {Corcella}, \citenamefont {Hektor}, \citenamefont
  {Hutsi}, \citenamefont {Kadastik}, \citenamefont {Panci}, \citenamefont
  {Raidal}, \citenamefont {Sala},\ and\ \citenamefont
  {Strumia}}]{Cirelli:2010xx}%
  \BibitemOpen
  \bibfield  {author} {\bibinfo {author} {\bibfnamefont {M.}~\bibnamefont
  {Cirelli}}, \bibinfo {author} {\bibfnamefont {G.}~\bibnamefont {Corcella}},
  \bibinfo {author} {\bibfnamefont {A.}~\bibnamefont {Hektor}}, \bibinfo
  {author} {\bibfnamefont {G.}~\bibnamefont {Hutsi}}, \bibinfo {author}
  {\bibfnamefont {M.}~\bibnamefont {Kadastik}}, \bibinfo {author}
  {\bibfnamefont {P.}~\bibnamefont {Panci}}, \bibinfo {author} {\bibfnamefont
  {M.}~\bibnamefont {Raidal}}, \bibinfo {author} {\bibfnamefont
  {F.}~\bibnamefont {Sala}}, \ and\ \bibinfo {author} {\bibfnamefont
  {A.}~\bibnamefont {Strumia}},\ }\href {\doibase
  10.1088/1475-7516/2012/10/E01, 10.1088/1475-7516/2011/03/051} {\bibfield
  {journal} {\bibinfo  {journal} {JCAP}\ }\textbf {\bibinfo {volume} {1103}},\
  \bibinfo {pages} {051} (\bibinfo {year} {2011})},\ \bibinfo {note} {[Erratum:
  JCAP1210,E01(2012)]},\ \Eprint {http://arxiv.org/abs/1012.4515}
  {arXiv:1012.4515 [hep-ph]} \BibitemShut {NoStop}%
\bibitem [{\citenamefont {James}\ and\ \citenamefont
  {Roos}(1975)}]{James:1975dr}%
  \BibitemOpen
  \bibfield  {author} {\bibinfo {author} {\bibfnamefont {F.}~\bibnamefont
  {James}}\ and\ \bibinfo {author} {\bibfnamefont {M.}~\bibnamefont {Roos}},\
  }\href {\doibase 10.1016/0010-4655(75)90039-9} {\bibfield  {journal}
  {\bibinfo  {journal} {Comput. Phys. Commun.}\ }\textbf {\bibinfo {volume}
  {10}},\ \bibinfo {pages} {343} (\bibinfo {year} {1975})}\BibitemShut
  {NoStop}%
\bibitem [{\citenamefont {Aaboud}\ \emph
  {et~al.}(2018{\natexlab{a}})\citenamefont {Aaboud} \emph
  {et~al.}}]{Aaboud:2017phn}%
  \BibitemOpen
  \bibfield  {author} {\bibinfo {author} {\bibfnamefont {M.}~\bibnamefont
  {Aaboud}} \emph {et~al.} (\bibinfo {collaboration} {ATLAS}),\ }\href
  {\doibase 10.1007/JHEP01(2018)126} {\bibfield  {journal} {\bibinfo  {journal}
  {JHEP}\ }\textbf {\bibinfo {volume} {01}},\ \bibinfo {pages} {126} (\bibinfo
  {year} {2018}{\natexlab{a}})},\ \Eprint {http://arxiv.org/abs/1711.03301}
  {arXiv:1711.03301 [hep-ex]} \BibitemShut {NoStop}%
\bibitem [{\citenamefont {Sirunyan}\ \emph {et~al.}(2018)\citenamefont
  {Sirunyan} \emph {et~al.}}]{Sirunyan:2017jix}%
  \BibitemOpen
  \bibfield  {author} {\bibinfo {author} {\bibfnamefont {A.~M.}\ \bibnamefont
  {Sirunyan}} \emph {et~al.} (\bibinfo {collaboration} {CMS}),\ }\href
  {\doibase 10.1103/PhysRevD.97.092005} {\bibfield  {journal} {\bibinfo
  {journal} {Phys. Rev.}\ }\textbf {\bibinfo {volume} {D97}},\ \bibinfo {pages}
  {092005} (\bibinfo {year} {2018})},\ \Eprint
  {http://arxiv.org/abs/1712.02345} {arXiv:1712.02345 [hep-ex]} \BibitemShut
  {NoStop}%
\bibitem [{\citenamefont {Aaboud}\ \emph
  {et~al.}(2018{\natexlab{b}})\citenamefont {Aaboud} \emph
  {et~al.}}]{Aaboud:2017rzf}%
  \BibitemOpen
  \bibfield  {author} {\bibinfo {author} {\bibfnamefont {M.}~\bibnamefont
  {Aaboud}} \emph {et~al.} (\bibinfo {collaboration} {ATLAS}),\ }\href
  {\doibase 10.1140/epjc/s10052-017-5486-1} {\bibfield  {journal} {\bibinfo
  {journal} {Eur. Phys. J.}\ }\textbf {\bibinfo {volume} {C78}},\ \bibinfo
  {pages} {18} (\bibinfo {year} {2018}{\natexlab{b}})},\ \Eprint
  {http://arxiv.org/abs/1710.11412} {arXiv:1710.11412 [hep-ex]} \BibitemShut
  {NoStop}%
\bibitem [{\citenamefont {Aaboud}\ \emph
  {et~al.}(2018{\natexlab{c}})\citenamefont {Aaboud} \emph
  {et~al.}}]{Aaboud:2017aeu}%
  \BibitemOpen
  \bibfield  {author} {\bibinfo {author} {\bibfnamefont {M.}~\bibnamefont
  {Aaboud}} \emph {et~al.} (\bibinfo {collaboration} {ATLAS}),\ }\href
  {\doibase 10.1007/JHEP06(2018)108} {\bibfield  {journal} {\bibinfo  {journal}
  {JHEP}\ }\textbf {\bibinfo {volume} {06}},\ \bibinfo {pages} {108} (\bibinfo
  {year} {2018}{\natexlab{c}})},\ \Eprint {http://arxiv.org/abs/1711.11520}
  {arXiv:1711.11520 [hep-ex]} \BibitemShut {NoStop}%
\bibitem [{\citenamefont {Sirunyan}\ \emph {et~al.}(2017)\citenamefont
  {Sirunyan} \emph {et~al.}}]{Sirunyan:2017xgm}%
  \BibitemOpen
  \bibfield  {author} {\bibinfo {author} {\bibfnamefont {A.~M.}\ \bibnamefont
  {Sirunyan}} \emph {et~al.} (\bibinfo {collaboration} {CMS}),\ }\href
  {\doibase 10.1140/epjc/s10052-017-5317-4} {\bibfield  {journal} {\bibinfo
  {journal} {Eur. Phys. J.}\ }\textbf {\bibinfo {volume} {C77}},\ \bibinfo
  {pages} {845} (\bibinfo {year} {2017})},\ \Eprint
  {http://arxiv.org/abs/1706.02581} {arXiv:1706.02581 [hep-ex]} \BibitemShut
  {NoStop}%
\bibitem [{\citenamefont {Aaboud}\ \emph {et~al.}(2019)\citenamefont {Aaboud}
  \emph {et~al.}}]{Aaboud:2019yqu}%
  \BibitemOpen
  \bibfield  {author} {\bibinfo {author} {\bibfnamefont {M.}~\bibnamefont
  {Aaboud}} \emph {et~al.} (\bibinfo {collaboration} {ATLAS}),\ }\href@noop {}
  {\  (\bibinfo {year} {2019})},\ \Eprint {http://arxiv.org/abs/1903.01400}
  {arXiv:1903.01400 [hep-ex]} \BibitemShut {NoStop}%
\bibitem [{\citenamefont {Anand}\ \emph {et~al.}(2014)\citenamefont {Anand},
  \citenamefont {Fitzpatrick},\ and\ \citenamefont {Haxton}}]{Anand:2013yka}%
  \BibitemOpen
  \bibfield  {author} {\bibinfo {author} {\bibfnamefont {N.}~\bibnamefont
  {Anand}}, \bibinfo {author} {\bibfnamefont {A.~L.}\ \bibnamefont
  {Fitzpatrick}}, \ and\ \bibinfo {author} {\bibfnamefont {W.~C.}\ \bibnamefont
  {Haxton}},\ }\href {\doibase 10.1103/PhysRevC.89.065501} {\bibfield
  {journal} {\bibinfo  {journal} {Phys. Rev.}\ }\textbf {\bibinfo {volume}
  {C89}},\ \bibinfo {pages} {065501} (\bibinfo {year} {2014})},\ \Eprint
  {http://arxiv.org/abs/1308.6288} {arXiv:1308.6288 [hep-ph]} \BibitemShut
  {NoStop}%
\bibitem [{\citenamefont {Aprile}\ \emph {et~al.}(2017)\citenamefont {Aprile}
  \emph {et~al.}}]{Aprile:2017iyp}%
  \BibitemOpen
  \bibfield  {author} {\bibinfo {author} {\bibfnamefont {E.}~\bibnamefont
  {Aprile}} \emph {et~al.} (\bibinfo {collaboration} {XENON}),\ }\href
  {\doibase 10.1103/PhysRevLett.119.181301} {\bibfield  {journal} {\bibinfo
  {journal} {Phys. Rev. Lett.}\ }\textbf {\bibinfo {volume} {119}},\ \bibinfo
  {pages} {181301} (\bibinfo {year} {2017})},\ \Eprint
  {http://arxiv.org/abs/1705.06655} {arXiv:1705.06655 [astro-ph.CO]}
  \BibitemShut {NoStop}%
\bibitem [{\citenamefont {Aprile}\ \emph {et~al.}(2018)\citenamefont {Aprile}
  \emph {et~al.}}]{Aprile:2018dbl}%
  \BibitemOpen
  \bibfield  {author} {\bibinfo {author} {\bibfnamefont {E.}~\bibnamefont
  {Aprile}} \emph {et~al.} (\bibinfo {collaboration} {XENON}),\ }\href
  {\doibase 10.1103/PhysRevLett.121.111302} {\bibfield  {journal} {\bibinfo
  {journal} {Phys. Rev. Lett.}\ }\textbf {\bibinfo {volume} {121}},\ \bibinfo
  {pages} {111302} (\bibinfo {year} {2018})},\ \Eprint
  {http://arxiv.org/abs/1805.12562} {arXiv:1805.12562 [astro-ph.CO]}
  \BibitemShut {NoStop}%
\end{thebibliography}%

\end{document}